\documentclass[12pt]{iopart}

\pdfoutput=1
\usepackage[colorlinks=true]{hyperref}
\usepackage{iopams}
\usepackage{amssymb}
\usepackage{color}
\usepackage{subfig}

\usepackage{graphicx}
\usepackage{oldgerm}
\usepackage{float}
\usepackage{verbatim}
\usepackage{epsfig}

\graphicspath{{images/}}

\def\be{\begin{equation}}
\def\ee{\end{equation}}

\usepackage{setspace}

\begin{document}

\bibliographystyle{unsrt}

\title[Optimal Latching in a Potts Model]{Optimal Region of Latching Activity in an Adaptive Potts Model for Networks of Neurons}
\author{Mohammad-Farshad Abdollah-nia, Mohammadkarim Saeedghalati and Abdolhossein Abbassian}
\address{School of Mathematics, Institute for Research in Fundamental Sciences (IPM), Tehran, Iran}
\eads{\mailto{m.f.abdollahnia@gmail.com}, \mailto{arsham@gmail.com} and \mailto{abbnet@mail.ipm.ir}}

\begin{abstract}
In statistical mechanics, the Potts model is a model for interacting spins with more than two discrete states. Neural networks which exhibit features of learning and associative memory can also be modeled by a system of Potts spins. A spontaneous behavior of hopping from one discrete attractor state to another (referred to as \emph{latching}) has been proposed to be associated with higher cognitive functions. Here we propose a model in which both the stochastic dynamics of Potts models and an adaptive potential function are present. A latching dynamics is observed in a limited region of the noise(temperature)-adaptation parameter space. We hence suggest noise as a fundamental factor in such alternations alongside adaptation. From a dynamical systems point of view, the noise-adaptation alternations may be the underlying mechanism for multi-stability in attractor based models. An optimality criterion for realistic models is finally inferred.
\end{abstract}


\maketitle

\section{Introduction}
Among statistical approaches to modeling neural networks, the Ising model, beside other binary models, has received a lot of attention as a maximum entropy pairwise model. An instance of such binary models is a Boltzmann machine which is a Monte Carlo version of  the Hopfield network. The Potts model \cite{Potts52} is essentially the generalization of the Ising model to more than two state network units and, like the Ising model, it first caught attention for its richness in physical applications \cite{Wu82}. Kanter was among the first who generalized the application of the Ising model in neural networks with features of learning and associative memory \cite{Amit85-2, Amit85-1} to Potts model \cite{Kanter88}. Some recent efforts have been dedicated to estimating the storage capacity of Potts model for associative memory \cite{Xiong10, Kropff05, Lowe07, Bolle93}.  Ising models constructed based on recorded data from cultured cortical neurons  have proven successful in providing a good description of the real data \cite{Schneidman06}. Although the quality and limitations of this model concerning pairwise correlations in larger networks are still under investigation \cite{Roudi09}, the Ising and Potts models are potentially capable of incorporating higher order correlations. Recently, these models with specific energy functions are found useful at many levels of image processing, including segmentation of an image into its constituent regions and multi-scale analysis of image data \cite{Mumford10}.

In their 2002 article, Hauser \etal suggested that a computational mechanism for recursion, which provides a capacity to generate an infinite range of expressions from a finite set of elements,  is the only uniquely human component of the faculty of language \cite{Hauser02}.  This argument, beside considerations about the local and global circuitry of the neocortex, is the basis of Treves and Roudi's  proposal for a Potts model with a hopping behavior among global network states, given the discrete nature of these attractor states. ``The trajectory \ldots will essentially include periods close to attracting states \ldots and rapid transitions between them. The system \emph{latches} between attractors'', as these authors describe it \cite{Treves05,ofTheEvol}. The dynamics of their model comprises sets of differential equations that determine the activation and adaptation behavior of network units \cite{ofTheEvol}. Other reports have studied the structure of latching transitions \cite{Kropff07-2, Russo08} as well as the issue of storing correlated patterns in such networks \cite{Kropff07-1}.

Interestingly, the latching problem in memory-based analyses bears a likeness to multi-stability problems, such as perceptual bi-stability: a phenomenon in which perception alternates between two distinct interpretations of an ambiguous stimulus. Moreno-Bote \etal challenge in their study the mainstream models that ascribe alternations between dominance of two or more competing neural populations to some form of slow adaptation acting on the dominant population, that leads to a switch in dominance to the competing population (\emph{oscillator models}). They propose noise as the main cause of alternations in their \emph{noise-driven attractor models} and construct a neurally plausible and experimentally consistent attractor model \cite{Moreno-Bote07}. There is a parallelism between the stochastic nature of dynamics in our model and noise in attractor models, as both models predict that alternations would cease in the hypothetical absence of noise: by \{eliminating noise/approaching zero temperature\} the system would \{settle down/freeze\} in one of the \{two percepts/several stored patterns\} and stay there indefinitely.

The model we present here is an alternative to the published approach by Treves \cite{Treves05}, with the major distinction of enjoying a stochastic dynamics traditionally present in physical Ising and Potts models. In fact, we have used the Markov chain Monte Carlo algorithm for a network with the Gibbs probability measure. Additionally, thanks to an adaptive potential function the network maintains the  adaptive quality of neuronal activity. The combination of these features results in a latching behavior, driven by both noise and adaptation with corresponding adjustable parameters--temperature and adaptation time constant, respectively. The latching we observe here is consistently qualified as a temporary retrieval of one stored pattern, followed by subsequent abandonment of that pattern and retrieval of another pattern.

In theory, given the two parameters of temperature (noise) and adaptation, it is not  evident at all how the latching behavior would be observed in different regions of the parameter space.  A key finding here (from simulations) is that this hopping behavior is limited to a particular region of adaptation versus noise, beyond which the system either locks in a specific attractor state, or disorderedly fluctuates over various configurations without any pattern retrieval at all. Even within the very area where latching behavior is observed, a privileged critical temperature ($T_\mathrm{c}=1$) inferred from statistical analysis suggests another preference, allowing us to distinguish an optimal region of activity. A comparison of the latching ``quality'' at such an optimal point with other sample points will also confirm our expectation of an optimal region.

The emergence of a sharply distinct region of activity is by and large nontrivial, and a theoretical description of various network states in terms of analytic solution to the dynamics equations in our stochastic multi-state (Potts) network might be a difficult task. Instead, we will endeavor in our current report to identify and demonstrate various network states using simulations of networks with various scales and characteristic parameters. We will establish the robustness of the observed latching region in networks of various size scales in terms of various order parameters;  examine the effect of simulation run time; corroborate the independence of the results from initial conditions and cue patterns;  study the interplay of noise and adaptation in the near-optimal region; and propose an optimality criterion and identify its region.

\section{Overview of model}
A \emph{Potts} network is a collection of $M$ interacting \emph{units}, each of which may be in one of multiple discrete states. It is actually a generalization of the \emph{Ising Model} with units having more than two possible states. A unit may represent a single neuron or a neural population, having multiple states of activity (action potential, firing rate, etc.) modeled as multiple Potts states.

In the model presented here, each unit may be in one of $S+1$ possible states\footnote{A more generalized (and realistic) condition is an inhomogeneous network in which $S$ might be different among units. We will not deal with such conditions here.}
\[
s \in \{0,\ldots,S\}
\]
consisting of $1$ ``null'' state ($s=0$), and $S$ ``genuine'' states ($s = 1, \ldots, S$). \footnote{Terminlogy borrowed from \cite{Treves05}.}

\subsection{Interaction of units}
The following energy function is defined for the network\footnote{For convenience, we omit the negative sign common in physical notations.}:
\begin{equation}\label{hamiltonian}
E= \frac{1}{2(S+1)^2}\sum_{i=1}^M{h_i^{s_i}}
\end{equation}
, where
\begin{equation}\label{h}
h_{i}^{s}=\sum_{j\neq i}\sum_{k,l=0}^{S} w_{ij}^{kl}u_{sk}u_{s_{j}l}
\end{equation}
describes the energy associated with unit $i$ being in an arbitrary state $s$, and $s_j$ denotes the \emph{current} state of the $j$th unit. $u_{sk}$ is defined based on  the following modification of the Kronecker's delta function is defined:
\begin{equation}\label{u}
u_{sk}=(S+1)\delta_{sk}-1
\end{equation}
which serves comparing two selected states of activity, $s$ and $k$. It assumes a value of $S$ if $s=k$, and $-1$ otherwise, thus the total summation over $k=0,\ldots,S$ adds up to zero. (Examine the case of $S=1$ -- the Ising model.)

There is also a \emph{weights matrix}, $w_{ij}^{kl}$, defined in section \ref{learningRule} which determines the relative preference of units $i$ and $j$ being in states $k$ and $l$, respectively.

\subsection{Learning rule}\label{learningRule}
A number of $p$ patterns are stored in the network with the weights matrix defined as follows:
\begin{equation}\label{w}
w_{ij}^{kl}=\frac{1}{(S+1)^2 M p}\sum_{\mu=1}^{p} u_{{\xi_i^\mu}k} u_{{\xi_j^\mu}l} (1-\delta_{k0})(1-\delta_{l0})
\end{equation}
in which $\xi_i^\mu$ represents the state of unit $i$ in pattern $\mu$. Notice that a weight of zero is associated with null states.

Substituting (\ref{w}) in (\ref{h}) shows that if a unit takes up a state which is defined in a stored pattern, the energy associated with that unit will be locally maximized. We will use this feature in section \ref{dynamics} to implement a higher rate of occurrence for our stored patterns via an appropriate distribution function.

\subsection{Dynamics}\label{dynamics}
To define a stochastic, while adaptive, dynamics for the system, we set the common Boltzmann rate, $e^{\beta h_i^s}$ ($\beta>0$), for the occurrence of state $s$ in unit $i$, and adaptively manipulate the ``attractiveness'' of a local attractor by virtually altering the energy function, $h_i^s$, based on the recent activity of each unit-state.

To accomplish this using a Monte-Carlo method of simulation, we randomly select a unit $i$ (which is in state $s_i$) in each iteration of the program, then choose a random state $r$ as a candidate for transition from $s_i$ to $r$. The transition occurs with the following probability (the Metropolis algorithm):
\begin{equation}\label{dynamics}
P(s_i \rightarrow r) =
 \cases{
 1 & if $\tilde{h}_i^r \geq \tilde{h}_i^{s_i}$ \cr
 \exp{\lbrack\beta(\tilde{h}_i^r-\tilde{h}_i^{s_i})\rbrack} & otherwise \cr
 }
\end{equation}
where
\begin{equation*}
\tilde{h}_i^k := h_i^k - {h^T}_i^k
\end{equation*}
represents an adapting potential, with $h_i^k$ coming from (\ref{h}) and ${h^T}_i^k$ being some adapting threshold with the following dynamics:
\begin{equation} \label{adapt}
\eqalign{
\tau\dot{h^T}_i^k &= u_{s_{i}k}-{h^T}_i^k \qquad k=1,\ldots,S \cr
{h^T}_i^0 &= 0 . \cr}
\end{equation}
Notice that there is no adaptation mechanism for null states.

The inverse of parameters $\tau$ (adaptation time constant) and $\beta$ (inverse temperature) represent the levels of \emph{noise} and \emph{adaptation} in the system respectively.

\section{Simulation and analysis}
Networks of various scales ($M=100$, $300$, $600$ and $900$) with $S=10$ were simulated over a domain of noise-adaptation combinations. Throughout this study, the value $S=10$ is used, unless otherwise stated. A number of $p\approx \frac{1}{30}M$ patterns were stored in each network. Patterns were generated following the method described in \cite{Treves05} which is capable of producing non- to highly-correlated patterns with desired levels of complexity and common units. In each pattern, a fraction of $a=0.5$ units were set to be in genuine states, with others being in null state. For the following studies, the correlation determinant factor ($\zeta$) was set to zero to produce uncorrelated patterns. For more details see the supplementary material.

\subsection{Overlaps behavior}
A primary quantity of interest, $O_\mu$, is the pattern retrieval reflected in the overlap (similarity) of the current state of the network with the stored pattern \(\mu\). It is simply measured by counting the number of common genuine unit-states between the current configuration of the network and each stored pattern, and then normalizing the result:
\begin{equation}\label{overlapsDef}
O_\mu = \frac{1}{M a}\sum_{i=1}^M \delta_{s_i \xi^\mu_i}(1-\delta_{0 s_i}).
\end{equation}

\begin{figure}
\begin{center}
\includegraphics{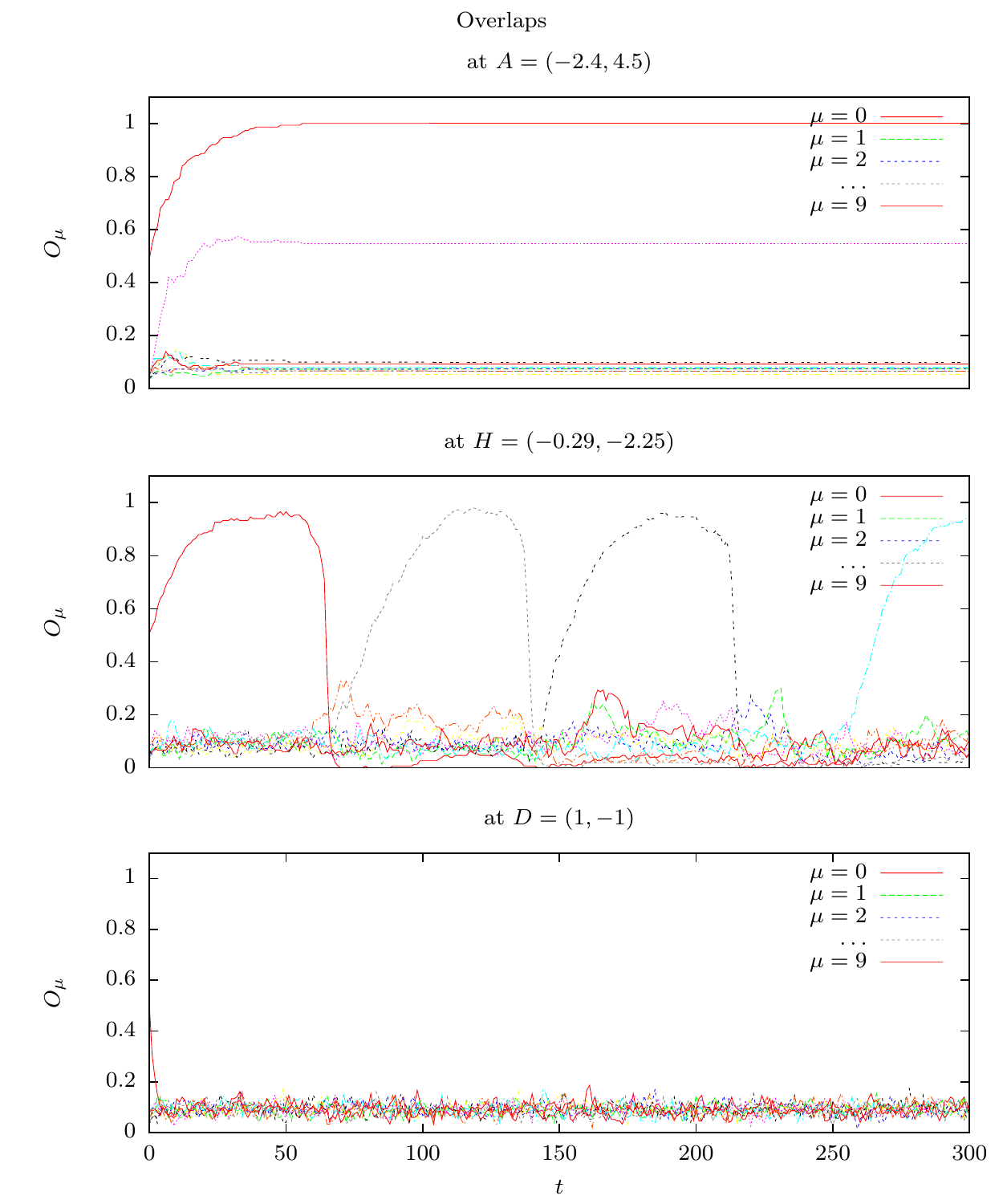}
\end{center}
\caption{Overlaps between the network status and the ten stored patterns in a network of $300$ units change over time. Notice that a pattern ($\mu=0$) is used as an initial cue in each run of the program. $P=(x, y)$ in each title means $\beta=10^{-x}$ and $\tau = 10^{-y}$. You can find the corresponding labels in figure~\ref{s10m300p10_OELand_top}. \label{probesAHD}}
\end{figure}

The resulting variations of overlaps, $O_\mu$, over time are depicted in figure~\ref{probesAHD} for three different pairs of $\beta$ and \(\tau\) selections. With proper selection of noise ($\beta^{-1}$) and adaptation ($\tau^{-1}$) parameters, a \emph{latching} behavior is observed in overlaps diagrams as the system hops from one retrieved pattern to another (figure~\ref{probesAHD}, middle.) Other types of behavior were also identified, in which the system is either underactive and frozen in a single pattern (figure~\ref{probesAHD}, top,) or overactive with no pattern retrieval (figure~\ref{probesAHD}, bottom.)

\subsection{Fluctuations landscape}
\begin{figure}
\begin{center}
\includegraphics{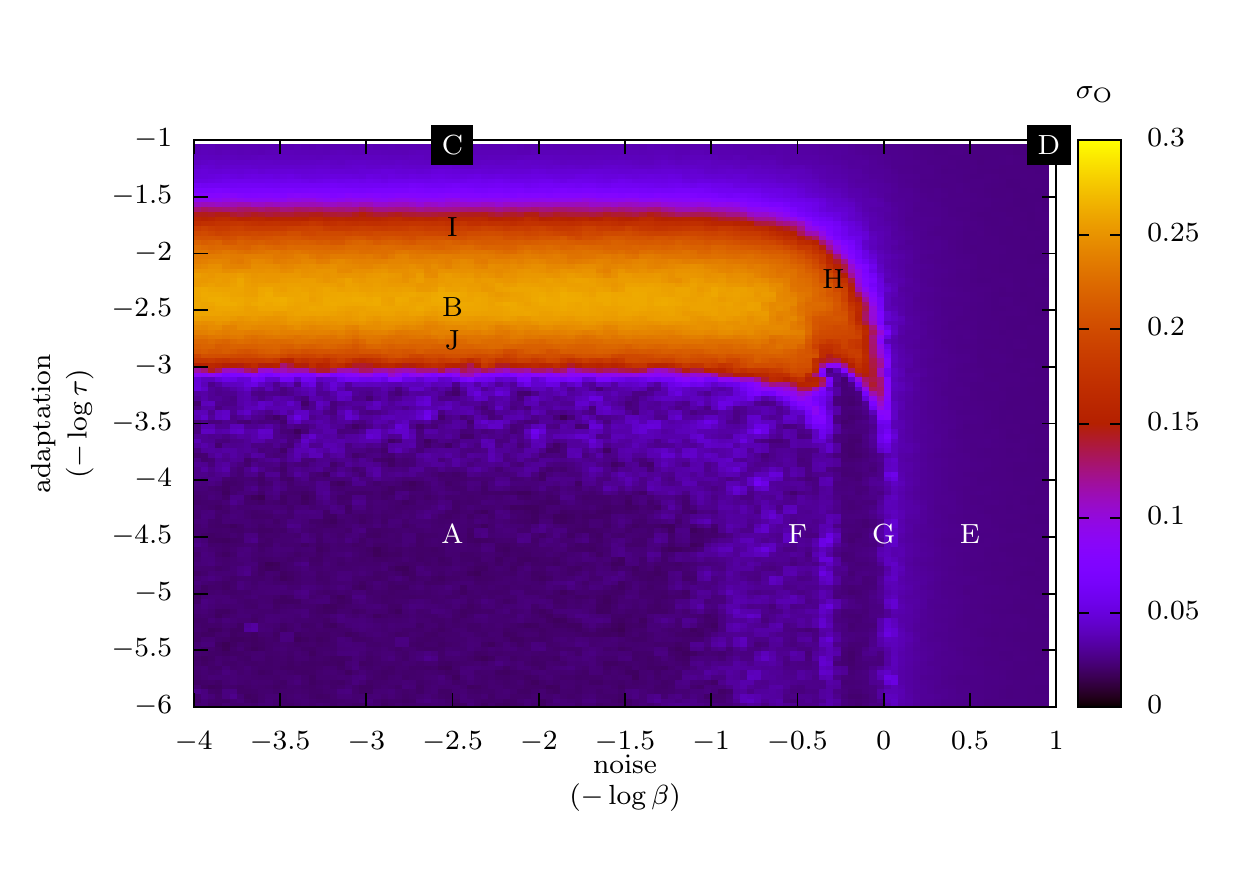}
\includegraphics{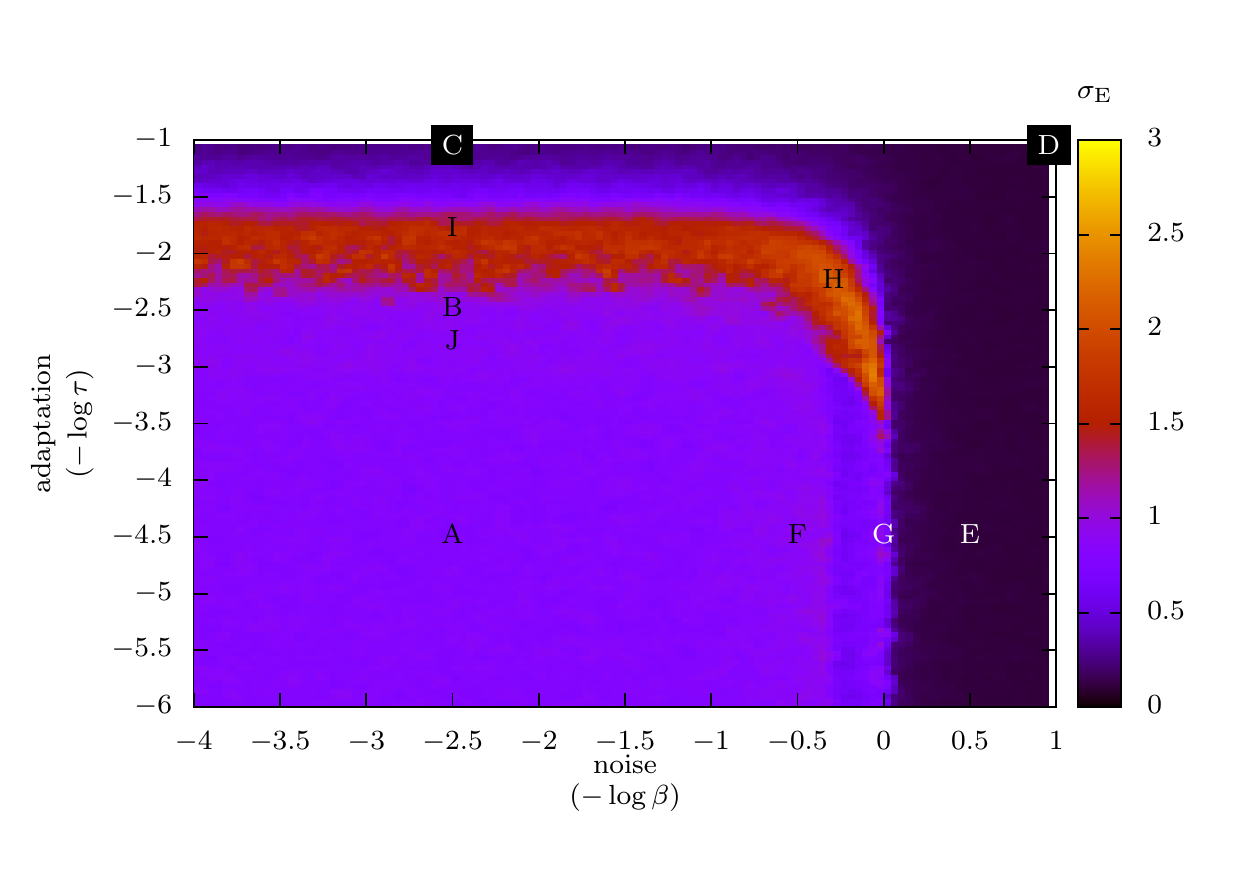}
\end{center}
\caption{Overlaps (top) and energy (bottom) fluctuations suggest a limited region of latching activity within the domain of noise ($-\log \beta$) and adaptation ($-\log \tau$) parameters. \label{s10m300p10_OELand_top}}
\end{figure}

To investigate the overall behavior of the network for each possible combination of noise and adaptation parameters, the averages of overlaps variations
\begin{equation*}\label{overlapsVariance}
{\sigma_O}^2 = \frac{1}{p}\sum_{\mu=1}^p (<{O_\mu}^2>_t - {<O_\mu>_t}^2)
\end{equation*}
 were measured over a wide grid of noise-adaptation sample points, where $<\dots>_t$ denotes averaging over a sufficiently long period of time at each point. The result is depicted in figure~\ref{s10m300p10_OELand_top} (top) for a network of size $M=300$ units, with $S=10$ for each unit.

Fluctuations of the total energy $E$ (see equation~(\ref{hamiltonian})), another order parameter, were also measured over the same grid points using the variance
\begin{equation*}
{\sigma_E}^2 = <E^2>_t - {<E>_t}^2 .
\end{equation*}
The result is plotted in figure~\ref{s10m300p10_OELand_top} (bottom.)

\begin{figure}
\begin{center}
\includegraphics{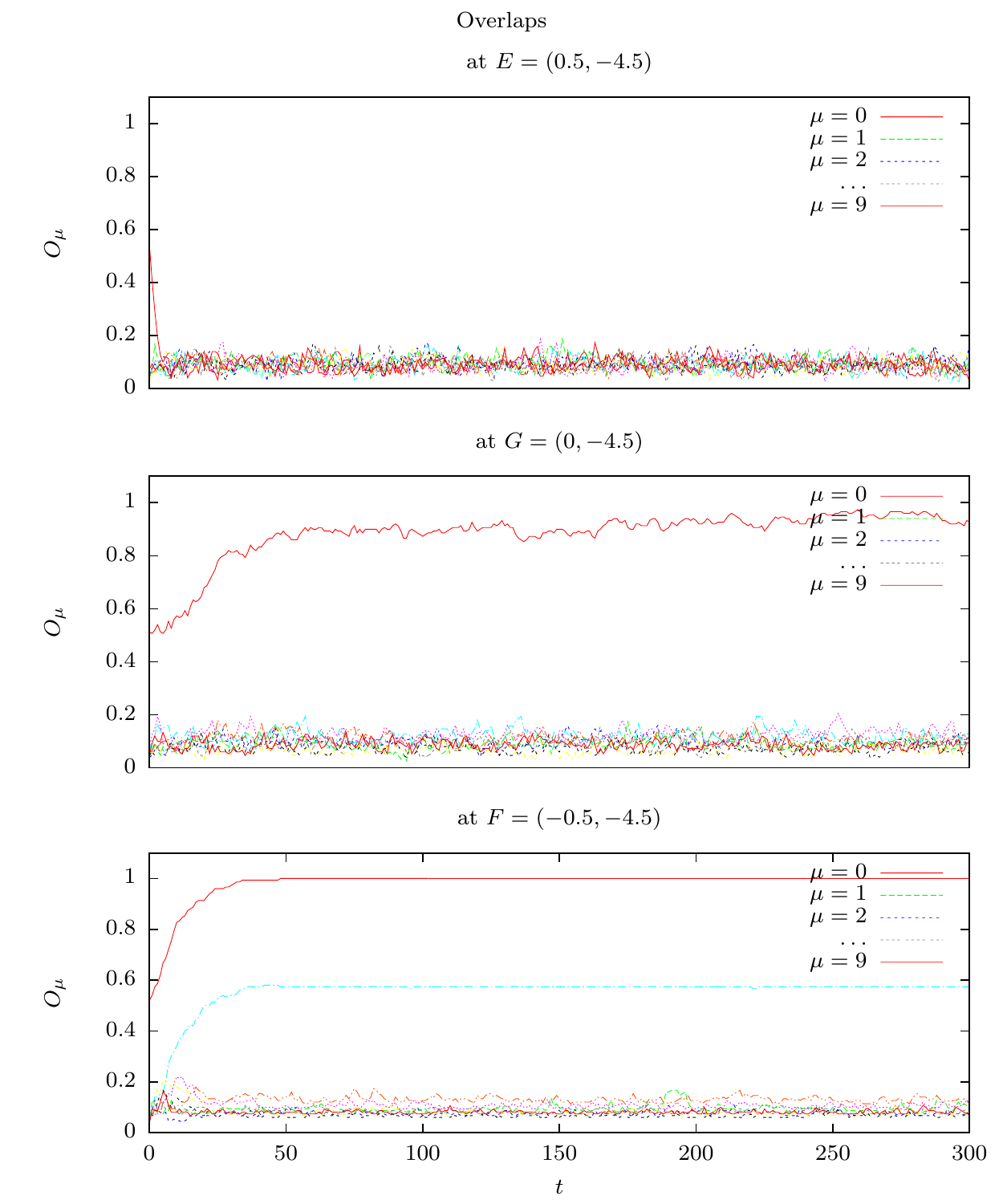}
\end{center}
\caption{Overlaps between the network status and the ten stored patterns in a network of $300$ units change over time. These graphs show three cases with a common, relatively slow adaptation. Noise, however, is different,  decreasing from the top panel to the bottom. At these points no latching behavior is observed. $P=(x, y)$ in each title means $\beta=10^{-x}$ and $\tau = 10^{-y}$. You can find the corresponding labels in figure~\ref{s10m300p10_OELand_top}.  \label{probesEFG}}
\end{figure}

As shown in figure~\ref{s10m300p10_OELand_top} the confined region of maximum fluctuations is the region that latching behavior occurs. Looking at sample points $A$,$H$ and $D$ studied in figure~\ref{probesAHD} confirms our expectation that the lower left section of the plot is in fact a \emph{frozen} region of activity if considered in a sufficiently short period of time compared to the adaptation time constant (see section~\ref{sectScaling}). The rest of the landscape belongs to an overactive or \emph{dead} region of pattern retrieval. At all three points $C$, $D$, and $E$, the behavior of overlaps diagram is similar, at least in appearance (see figures~\ref{probesAHD} and \ref{probesEFG}, the graph for point $C$ looks similar hence not shown for brevity). In this region, the system is too active in terms of unit-state fluctuations ($q_\mathrm{EA}$ parameter, section~\ref{sectScaling}) for any patterns to be retrieved, which means ironically dead in terms of pattern retrieval.

\begin{figure}
\begin{center}
\includegraphics{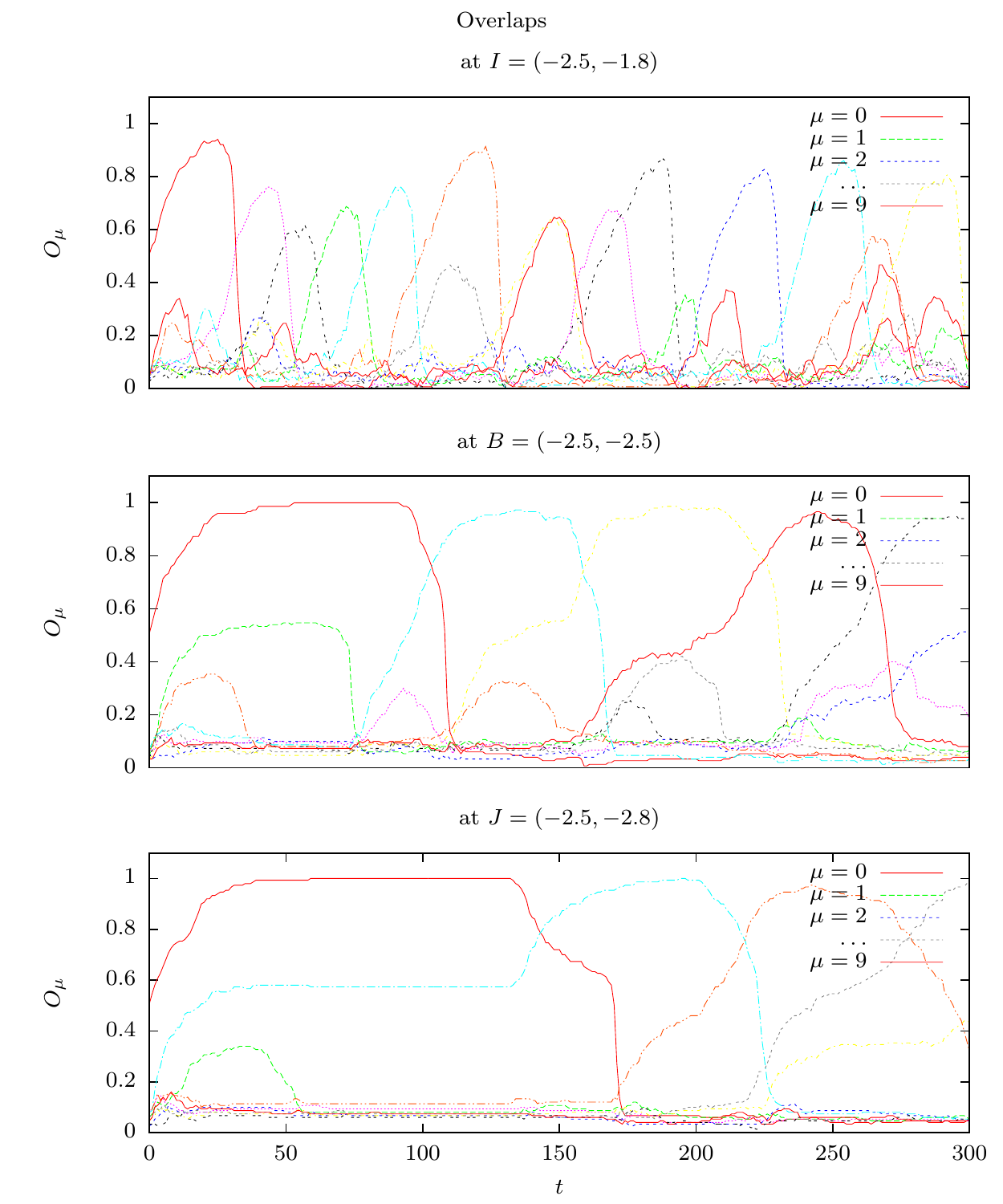}
\end{center}
\caption{Overlaps between the network status and the ten stored patterns in a network of $300$ units change over time. These graphs show three cases with a common, relatively low  noise value. Adaptation, though, is different,  decreasing from the top panel to the bottom. $P=(x, y)$ in each title means $\beta=10^{-x}$ and $\tau = 10^{-y}$. You can find the corresponding labels in figure~\ref{s10m300p10_OELand_top}.  \label{probesIBJ}}
\end{figure}

To reveal more details about the behavior of the network with various combinations of noise-adaptation parameters, several other sample points were labeled in figure~\ref{s10m300p10_OELand_top} and their overlaps graphs were sketched. The points were chosen to be cases with minimal noise, figures~\ref{probesIBJ}, or very slow adaptation, figure~\ref{probesEFG}.


\section{Size scaling, run time and initial conditions} \label{sectScaling}
\begin{figure}
\begin{center}
\subfloat[$\sigma_O$ for $-\log{\beta}=-0.25$]{\includegraphics[width=0.5\textwidth]{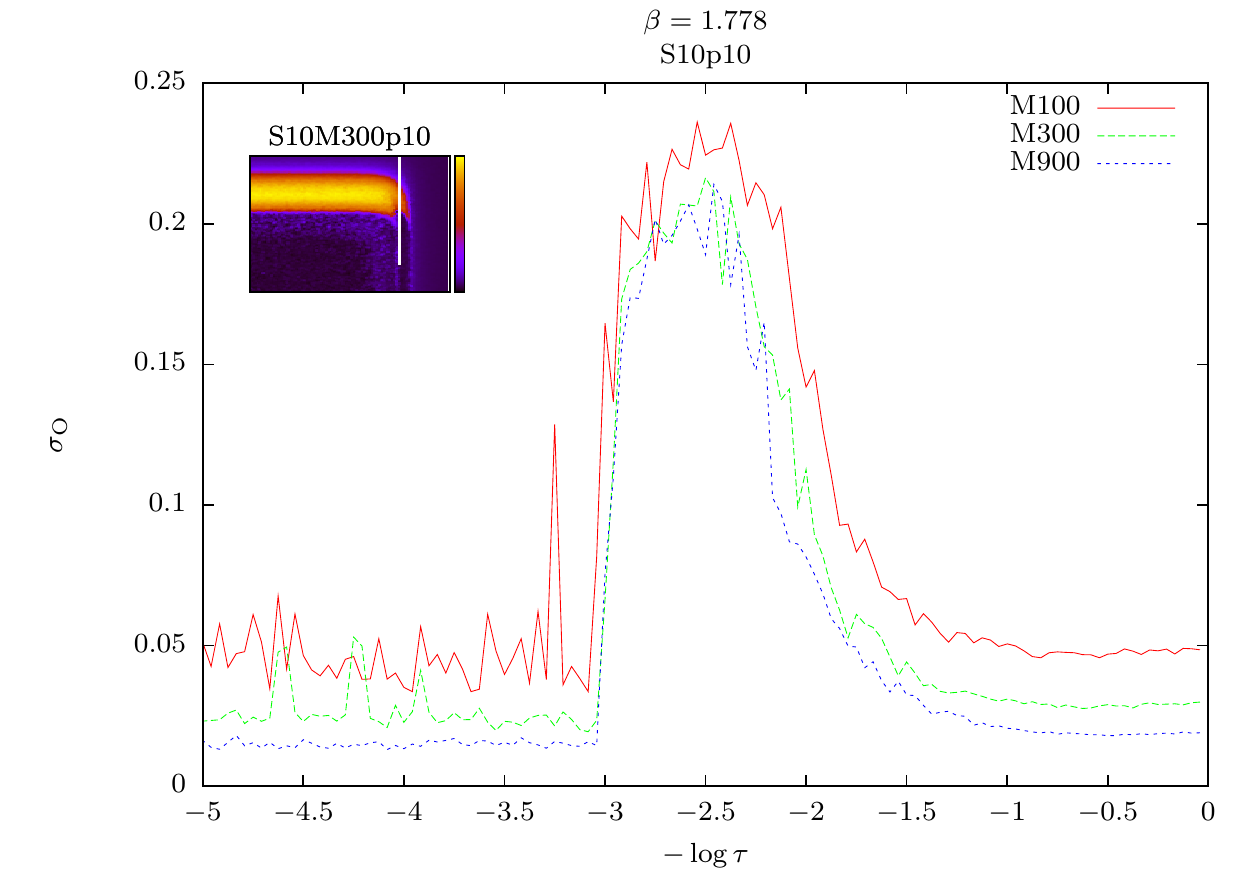}\label{beta1_7_O}}
\subfloat[$\sigma_O$ for $-\log{\tau}=-2.7$]{\includegraphics[width=0.5\textwidth]{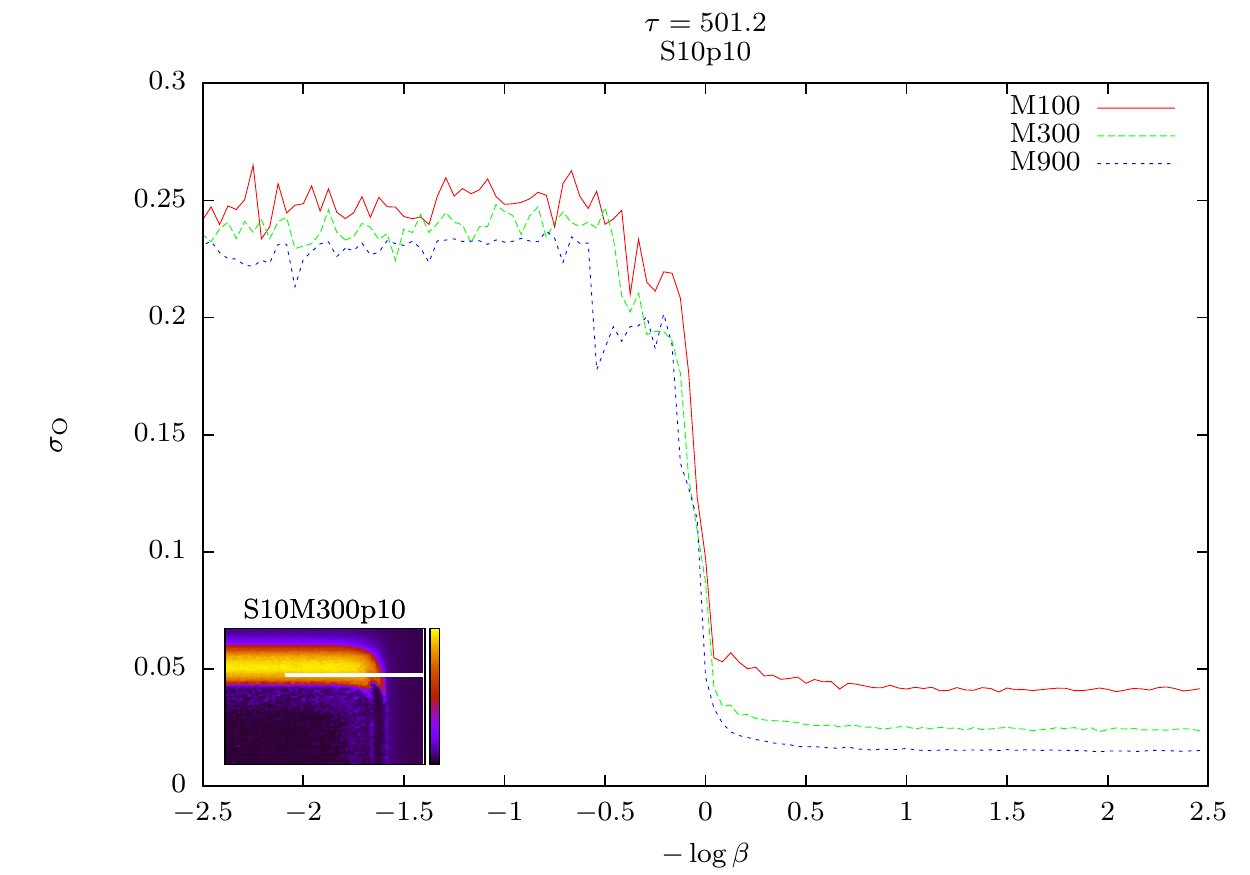}\label{tau501_O}}\\
\subfloat[$\sigma_E$ for $-\log{\beta}=-0.25$]{\includegraphics[width=0.5\textwidth]{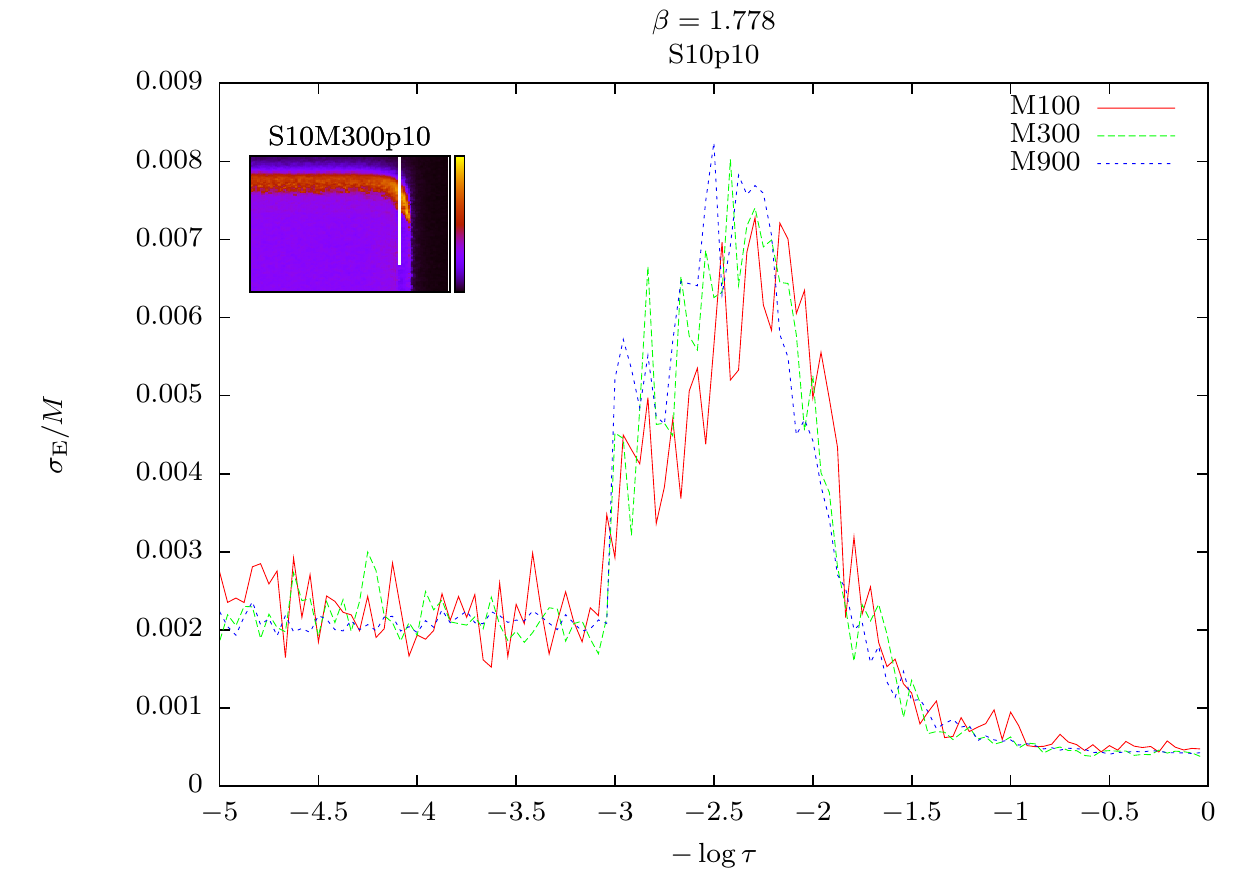}\label{beta1_7_E}}
\subfloat[$\sigma_E$ for $-\log{\tau}=-2.7$]{\includegraphics[width=0.5\textwidth]{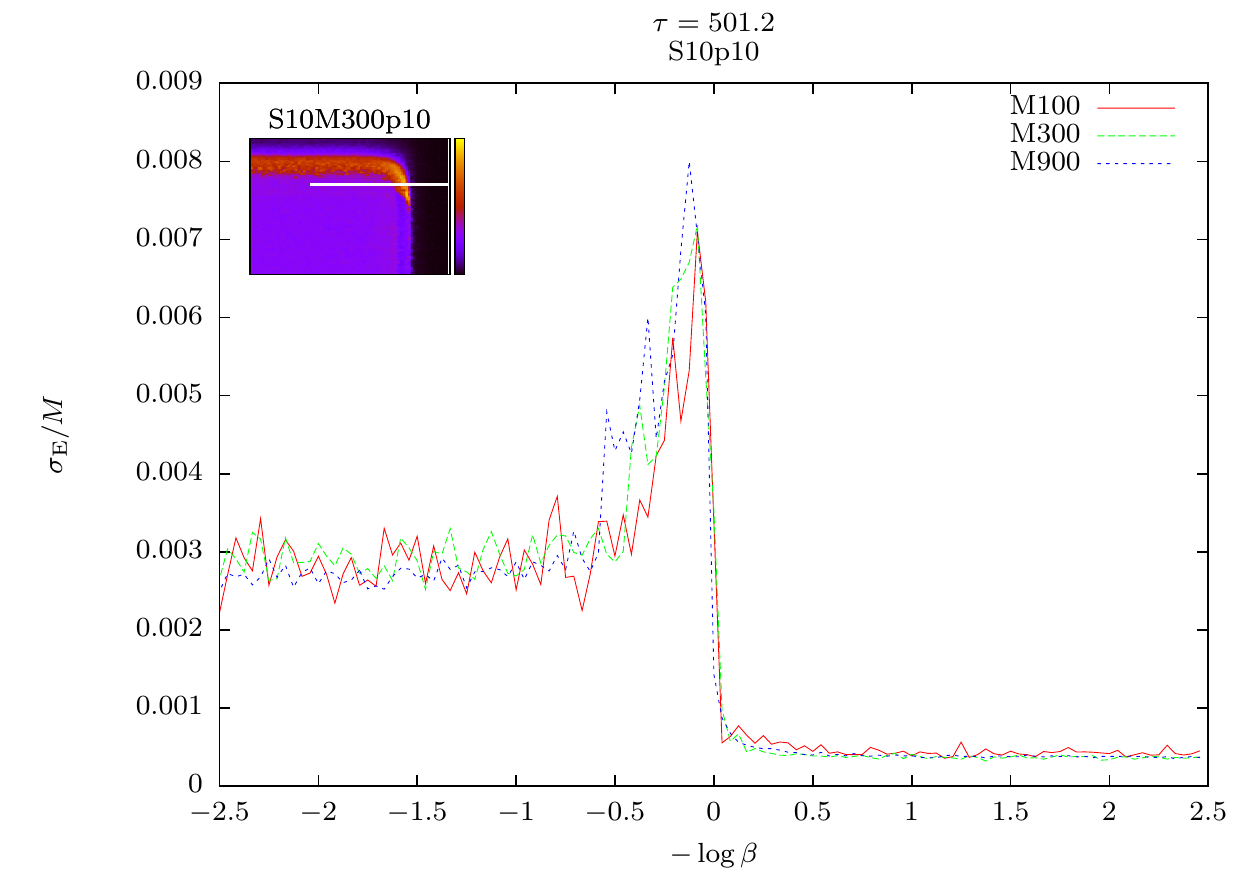}\label{tau501_E}}\\
\subfloat[$q_\mathrm{EA}$ for $-\log{\beta}=-1.5$]{\includegraphics[width=0.5\textwidth]{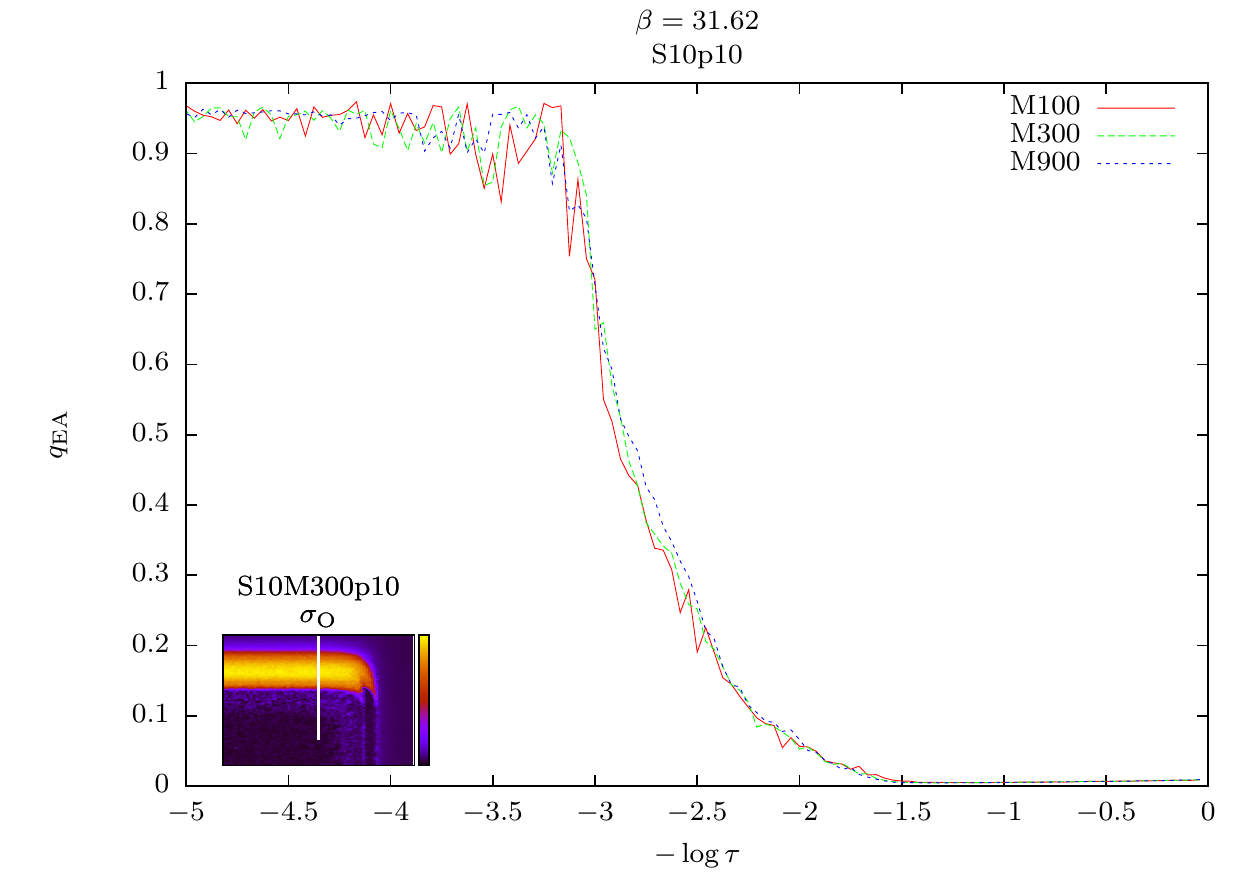}\label{beta31_6_q}}
\subfloat[$q_\mathrm{EA}$ for $-\log{\tau}=-1.8$]{\includegraphics[width=0.5\textwidth]{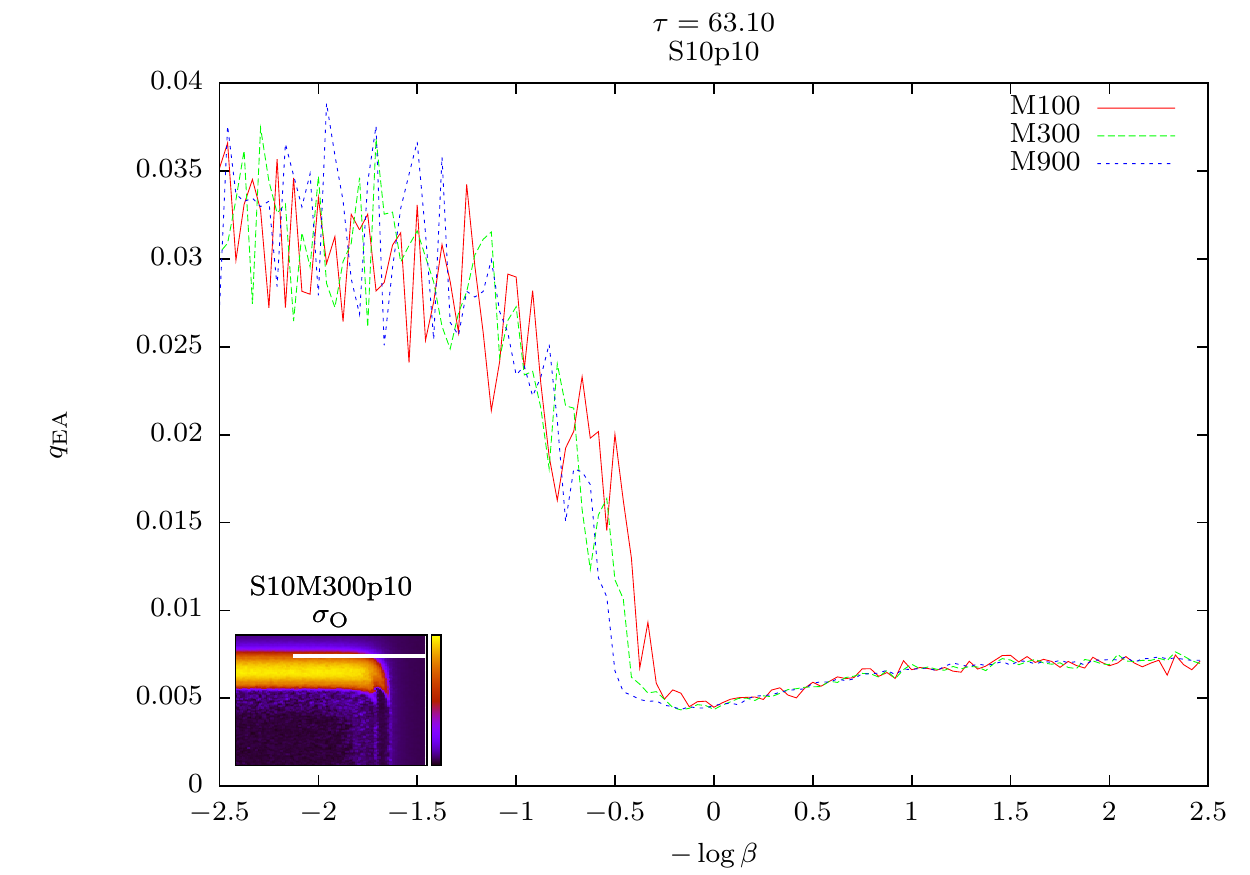}\label{tau63_q}}
\end{center}
\caption{Several order parameters in various sections of the noise-adaptation landscape (inset, see figure~\ref{s10m300p10_OELand_top}) are examined for different network sizes. Panels (a), (c), and (e) are vertical sections with a fixed value of $\beta$ (varying  adaptation). Panels (b), (d), and (f) are horizontal sections with a fixed adaptation (varying noise). A perfect consistency is observed. \label{scalings}}.
\end{figure}

The overall behavior of the system is invariant with respect to various network sizes: Several sections of figure~\ref{s10m300p10_OELand_top} were selected and replotted for different network sizes, $M = 100$, $300$, $900$. Some of these sections are depicted in figure~\ref{scalings}. The corresponding regions of activity evidently match in different size scales.

At this stage of the study, a third order parameter besides $\sigma_O$ and $\sigma_E$ was also examined, which provided a better understanding of the observed regions of activity. The Edwards-Anderson order parameter defined as
\begin{equation}\label{q_EA}
q_\mathrm{EA}=\frac{1}{MS(S+1)}\sum_{i,k}{{\left\langle u_{s_{i}k}\right\rangle}_t}^2
\end{equation}
is also plotted in figures~\ref{beta31_6_q} and \ref{tau63_q} for various sizes of the network in sections passing through different regions. The figures reveal that in the region of high $\sigma_\mathrm{O}$ and high $\sigma_\mathrm{E}$, the parameter $q_\mathrm{EA}$ varies gradually from its maximum to minimum value. To see what  $q_\mathrm{EA}$ measures, notice that $\sum_{k}{{\left\langle u_{s_{i}k}\right\rangle}_t}^2$ in equation~(\ref{q_EA}) is the average state of a unit $i$ in time, which takes the value $S^2+S$ if the unit is in a fixed state $s_i$, and vanishes if the unit is randomly and uniformly fluctuating between all states (cf the definition of $u_{sk}$ in~(\ref{u})). Averaging this for all units and normalizing such that the maximum value is $1$ yields equation~(\ref{q_EA}). This quantity is hence a better indicator of overall network activity as it clearly distinguishes between active and silent network states. Therefore, it is the high value of $q_\mathrm{EA}$ more than the low value of $\sigma_\mathrm{O}$ that indicates the frozen region.

\begin{figure}
\begin{center}
\subfloat[$\sigma_O$ for $t = 1500$ to $5000$]{\includegraphics[width=0.5\textwidth]{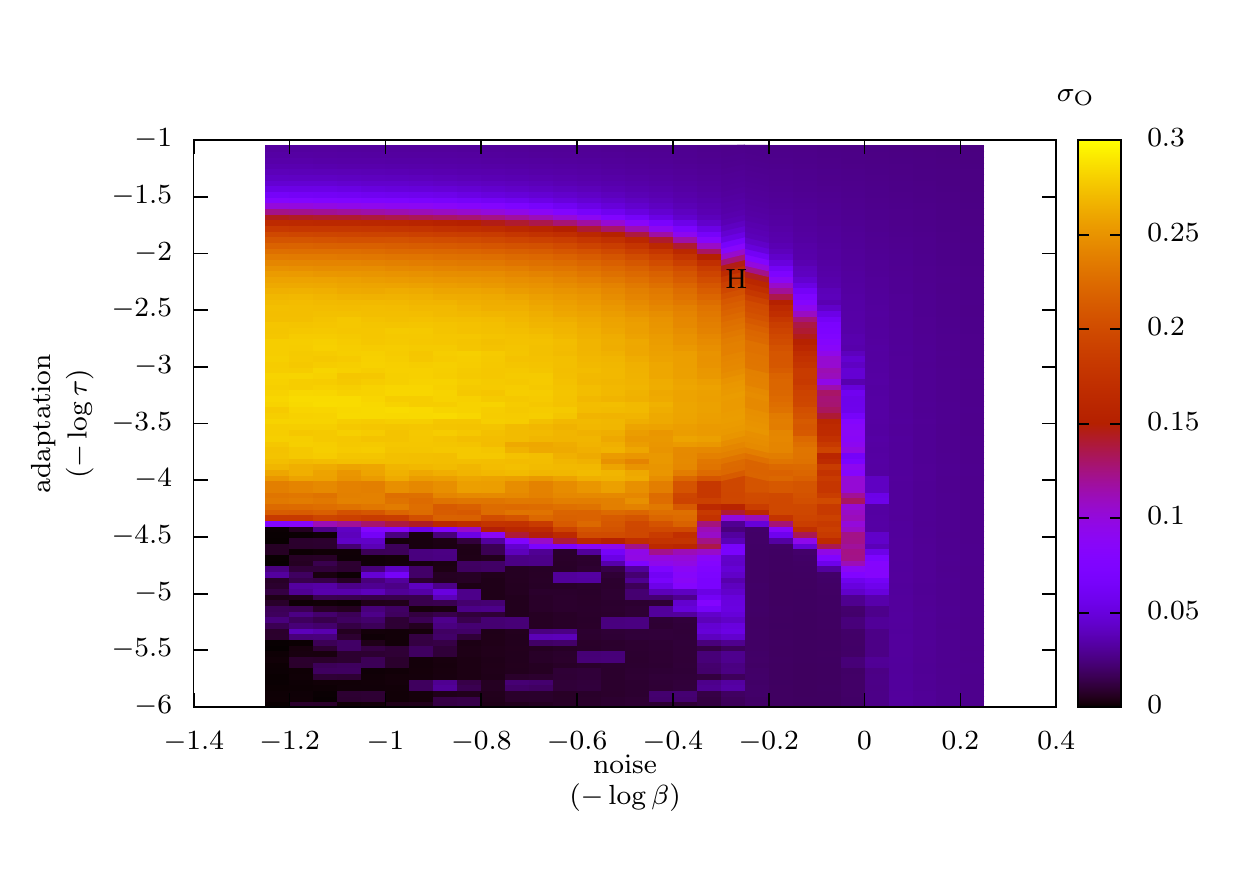}\label{longland_O}}
\subfloat[$\sigma_O$ for $t = 1$ to $5000$]{\includegraphics[width=0.5\textwidth]{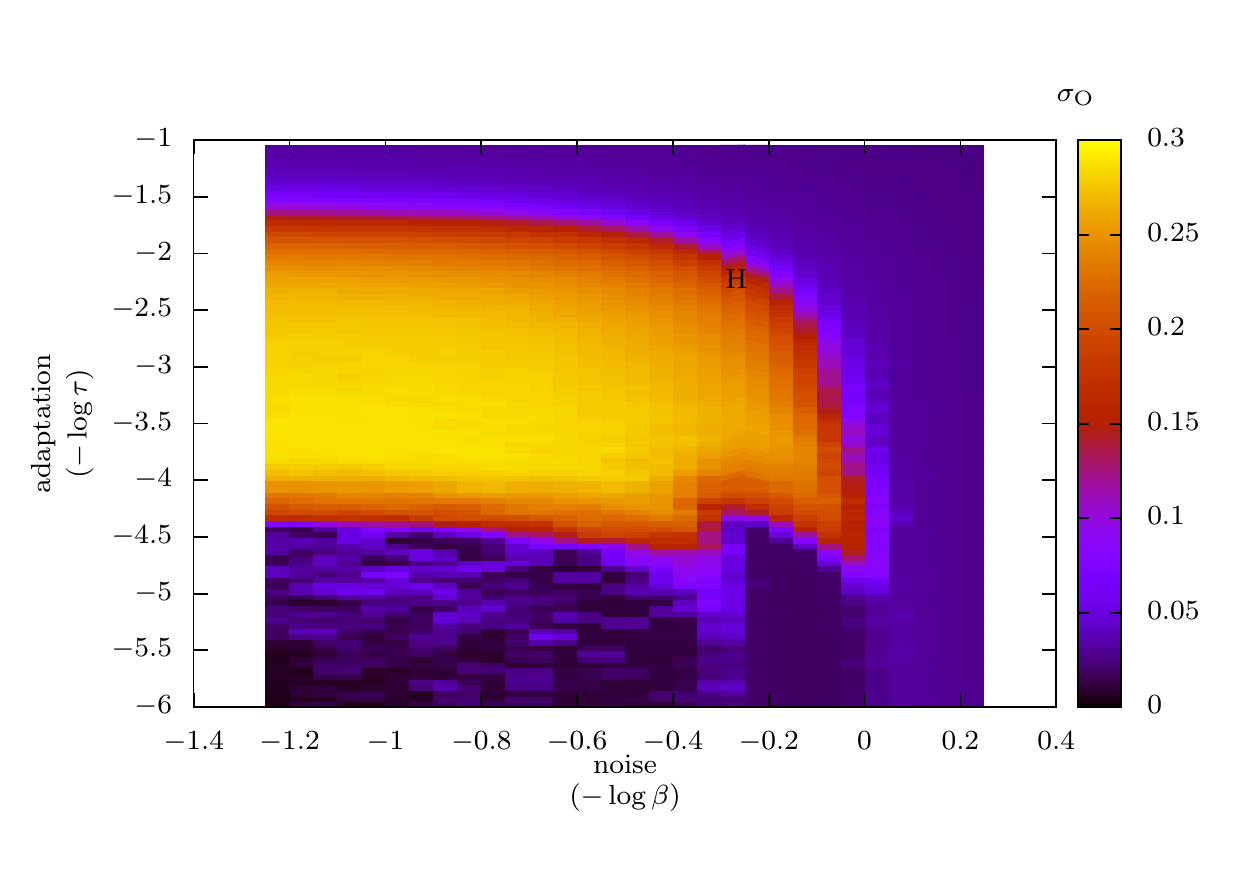}\label{longland_O_IK}}\\
\subfloat[$\sigma_E$ for $t = 1500$ to $5000$]{\includegraphics[width=0.5\textwidth]{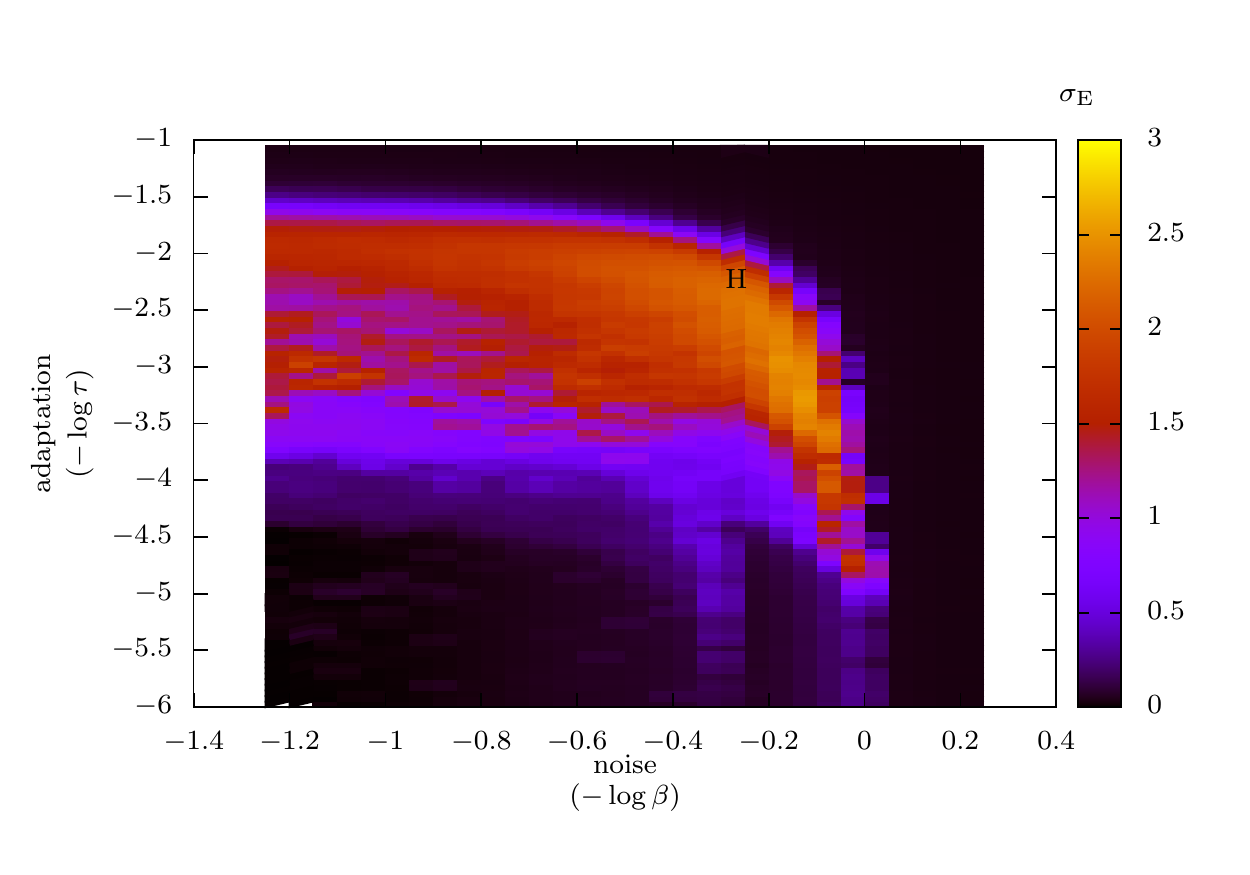}\label{longland_E}}
\subfloat[$\sigma_E$ for $t = 1$ to $5000$]{\includegraphics[width=0.5\textwidth]{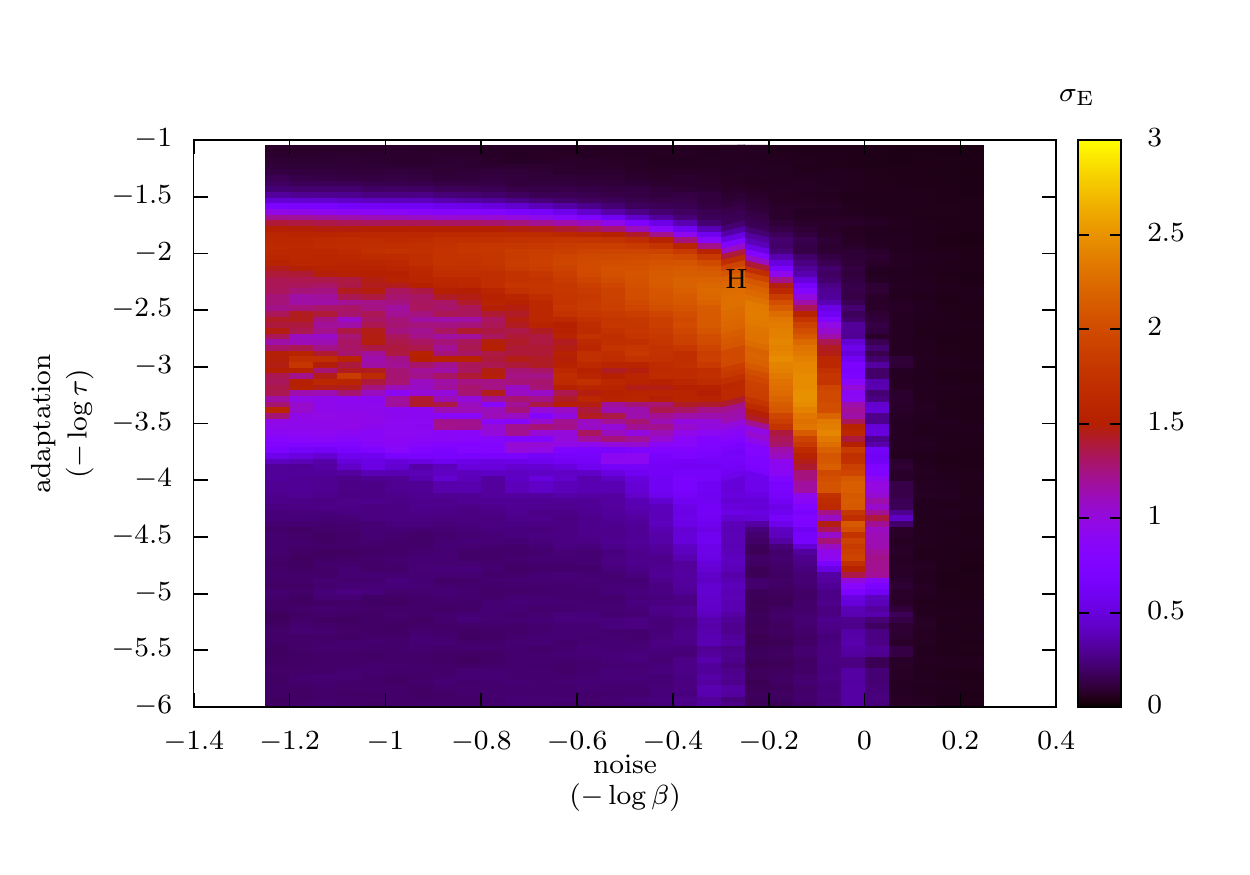}\label{longland_O_IK}}\\
\subfloat[$q_\mathrm{EA}$ for $t = 1500$ to $5000$]{\includegraphics[width=0.5\textwidth]{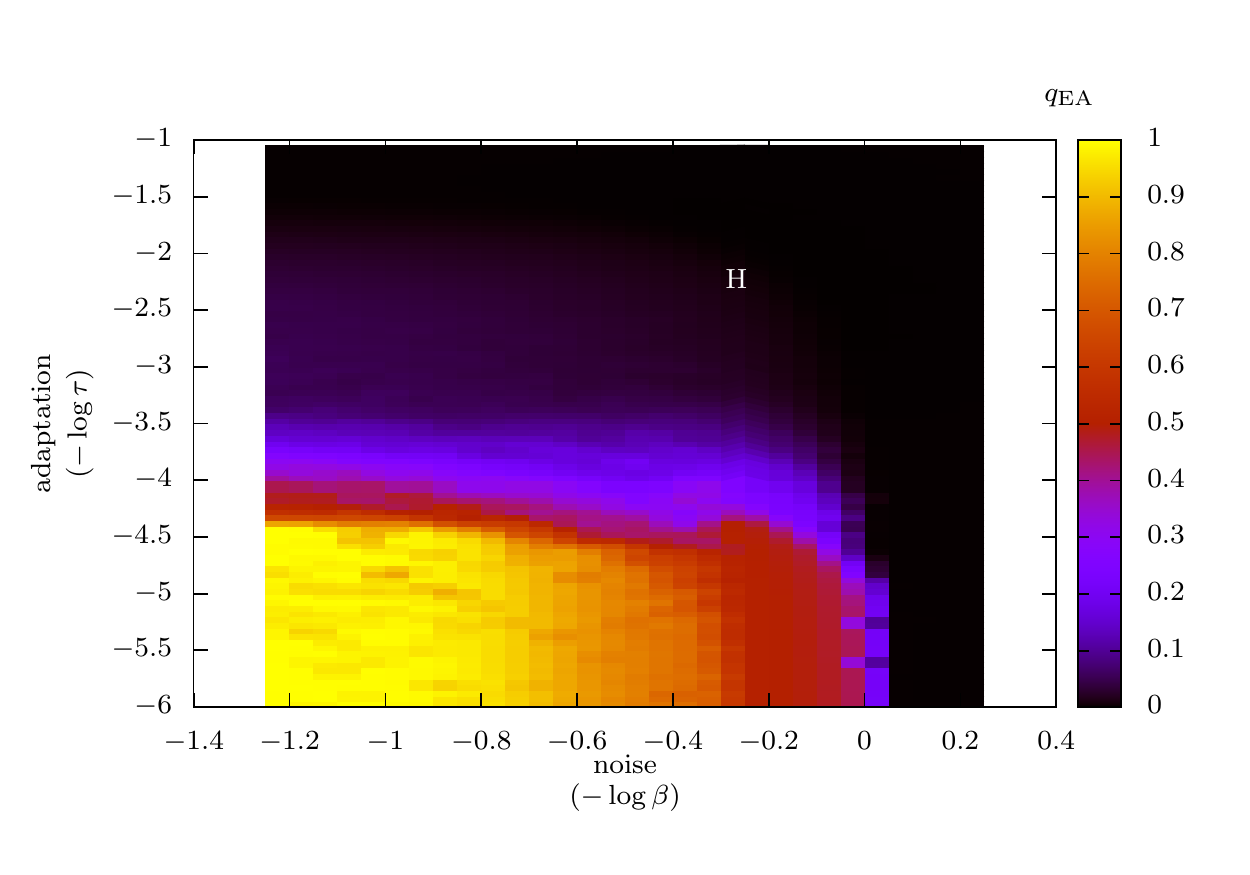}\label{longland_q}}
\subfloat[$q_\mathrm{EA}$ for $t = 1$ to $5000$]{\includegraphics[width=0.5\textwidth]{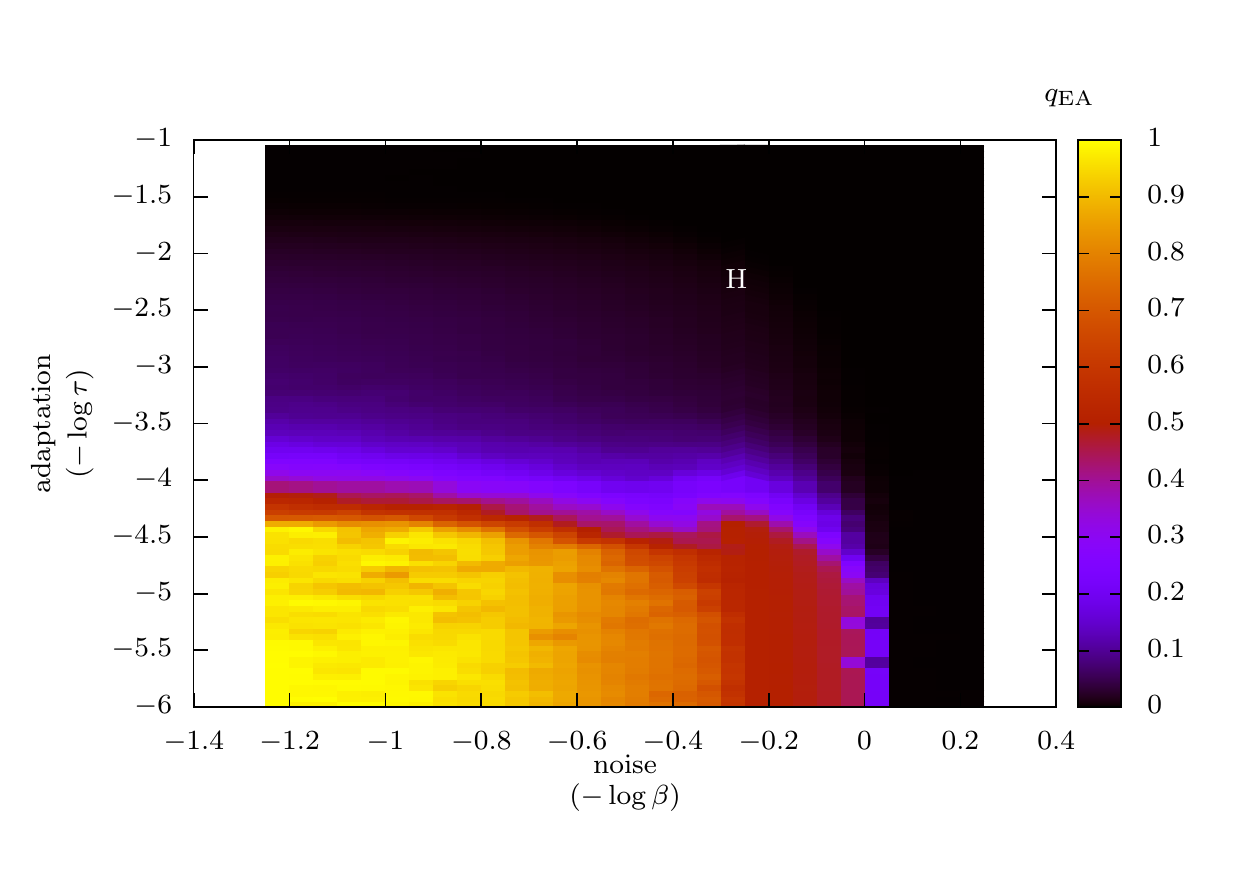}\label{longland_O_IK_IK}}
\end{center}
\caption{The noise-adaptation landscape for $5000$ steps. In the left panels, some initial portion of the data is ignored. Compare with figure~\ref{s10m300p10_OELand_top} where the run time is $300$ steps.  \label{longland}}
\end{figure}

Figures~\ref{beta31_6_q} and \ref{tau63_q} suggest that the shape and extent of the regions are robustly preserved under size scaling in the limited-sized networks that were studied. However, changing the simulation run time has a totally different effect on the extent of some regions. Figure~\ref{longland} shows a selection of the same landscape as figure~\ref{s10m300p10_OELand_top}, which is obtained through a much longer run time at each point. All the figures so far were obtained with a run time of $300$ ``steps,'' with each step here being $M$ single iterations of the program. Figure~\ref{longland}, however, is the result of $5000$ steps at each point. In the left panels, the data from the initial $1500$ steps of the simulation was ignored. The right panels where created using the full $5000$ steps. Not much difference is observed. The landscape view of $q_\mathrm{EA}$ was also plotted this time.

A noticeable difference between the short and the long simulation runs is observed. In the long runs, the region in which overlap fluctuations is observed extends more downwards, towards slower adaptations, or longer adaptation time constant. In other words, as we decrease the adaptation, the effect of small adaptation is still significant in the total overlap and energy fluctuations, because, although the retrieved patterns last for longer times, they finally switch to other patterns (non-optimal latching). The value of $q_\mathrm{EA}$ also increases more gradually in the long runs.

Consequently, one can argue that in reality there are no distinct phases or phase transitions if we look at the system in a sufficiently long time window, just a dynamics that slows down as the adaptation slows down. However, the maximum actual specific time scale of a real neural system sets an upper limit to the adaptation time constant, above which the system may be ``effectively'' frozen, or overactive, depending on the noise value. Moreover, an increase occurring in all the order parameters begins at around $-\log{\tau} = -1.7$, which is independent of the run time and network size. This also may be considered as a phase change phenomena. In fact, at around this point the dynamics is extremely sensitive to $\tau$ variations. With slow enough adaptation the system will have enough time to fully retrieve patterns. However, the lower limit for adaptation time constant (upper limit for adaptation speed) is, once again, systematically determined. If $\tau$ gets too small, the life time of retrieved patterns tends towards a time ``step.'' This means the adaptation is so fast that some parts of a pattern de-adapt before the whole pattern is retrieved, thus giving no time for the attractor state to rise. This ``step'' is an intrinsic property of real systems.

To ensure that our results are independent of various initial conditions and cue patterns, a section of figure~\ref{s10m300p10_OELand_top} was reexamined using a run period of $9000$ steps, with various random initial conditions and cue patterns. We also threw away the data from the first $3000$ steps of the total $9000$ steps. Four different cue patterns with two random initial conditions for each pattern were tried at each point (8 trials). The resulting standard deviations are shown in figures~\ref{initsOE} and \ref{initsq} with error bars.  The inset graphs show the corresponding section of study in figure~\ref{s10m300p10_OELand_top}.

\begin{figure}
\begin{center}
\includegraphics{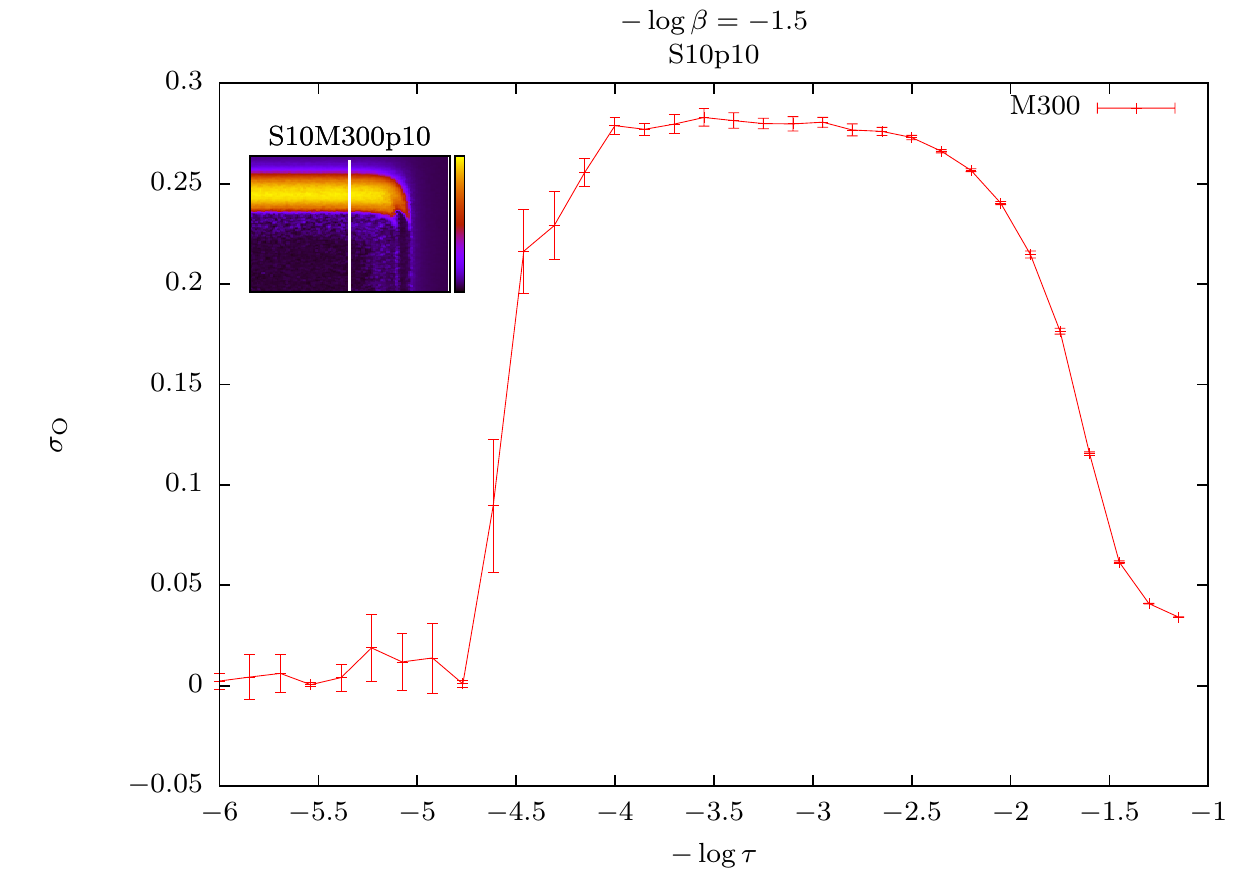}\\
\includegraphics{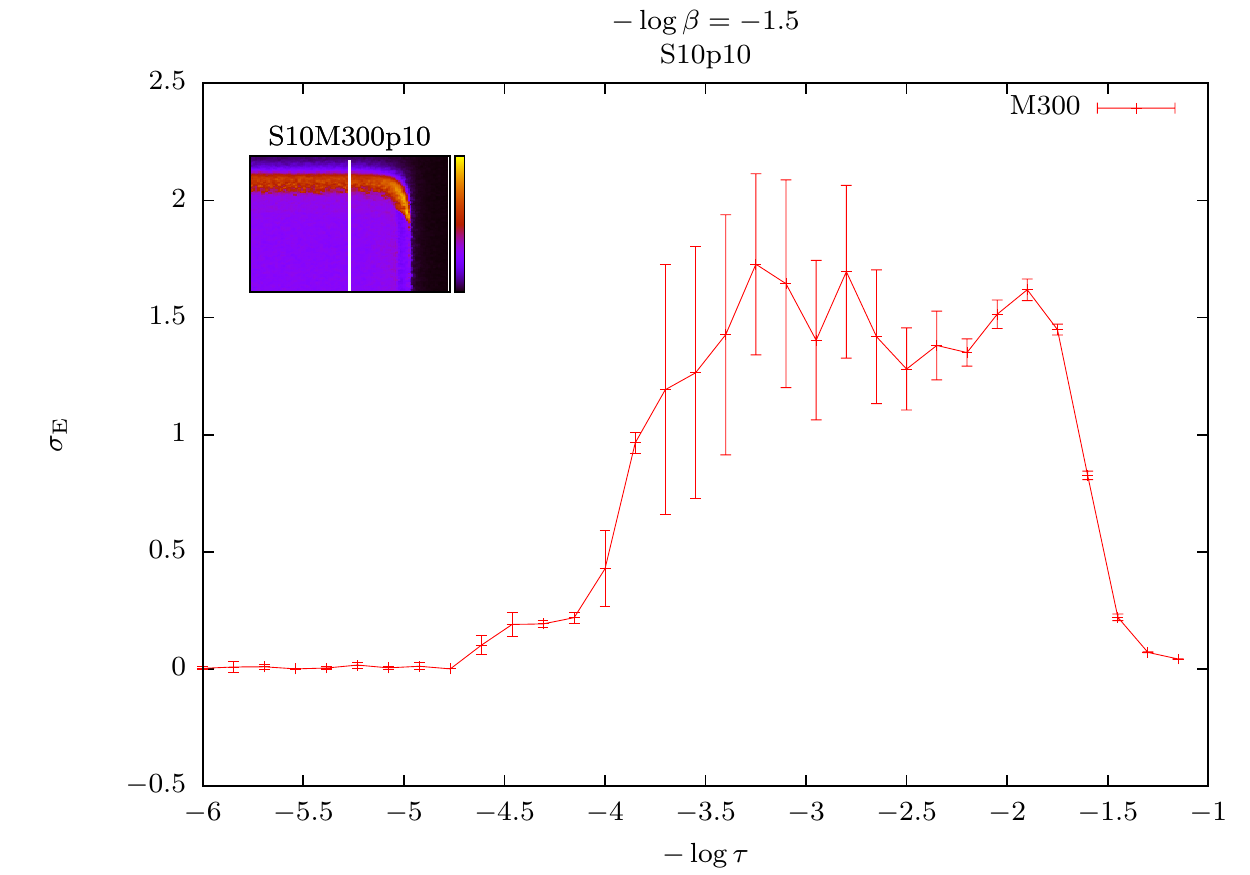}
\end{center}
\caption{Reliability check: deviations from mean values over various initial conditions
and various cue patterns are shown with error bars. The run period is $9000$ steps compared to 300 in figures~\ref{scalings} and \ref{s10m300p10_OELand_top} (the inset). Also, the initial $3000$ steps of each run is ignored. In these sections, the peak in the inset graph is extended more towards slower adaptations because in the longer runs, the non-optimal latching will still be observed with slower adaptations.
\label{initsOE}}
\end{figure}
\begin{figure}
\begin{center}
\includegraphics{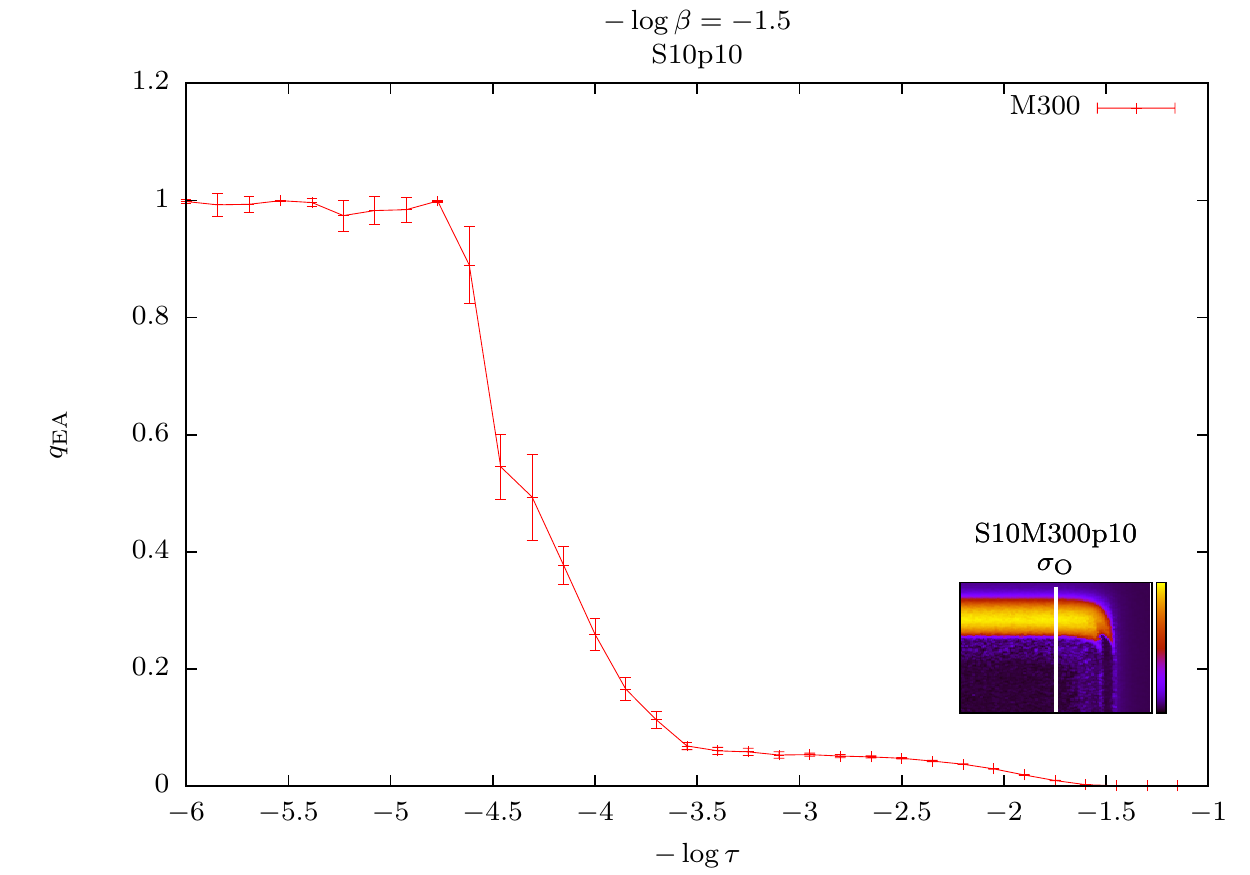}
\end{center}
\caption{Continued from figure~\ref{initsOE}.
\label{initsq}}
\end{figure}

\begin{figure}
\begin{center}
\includegraphics{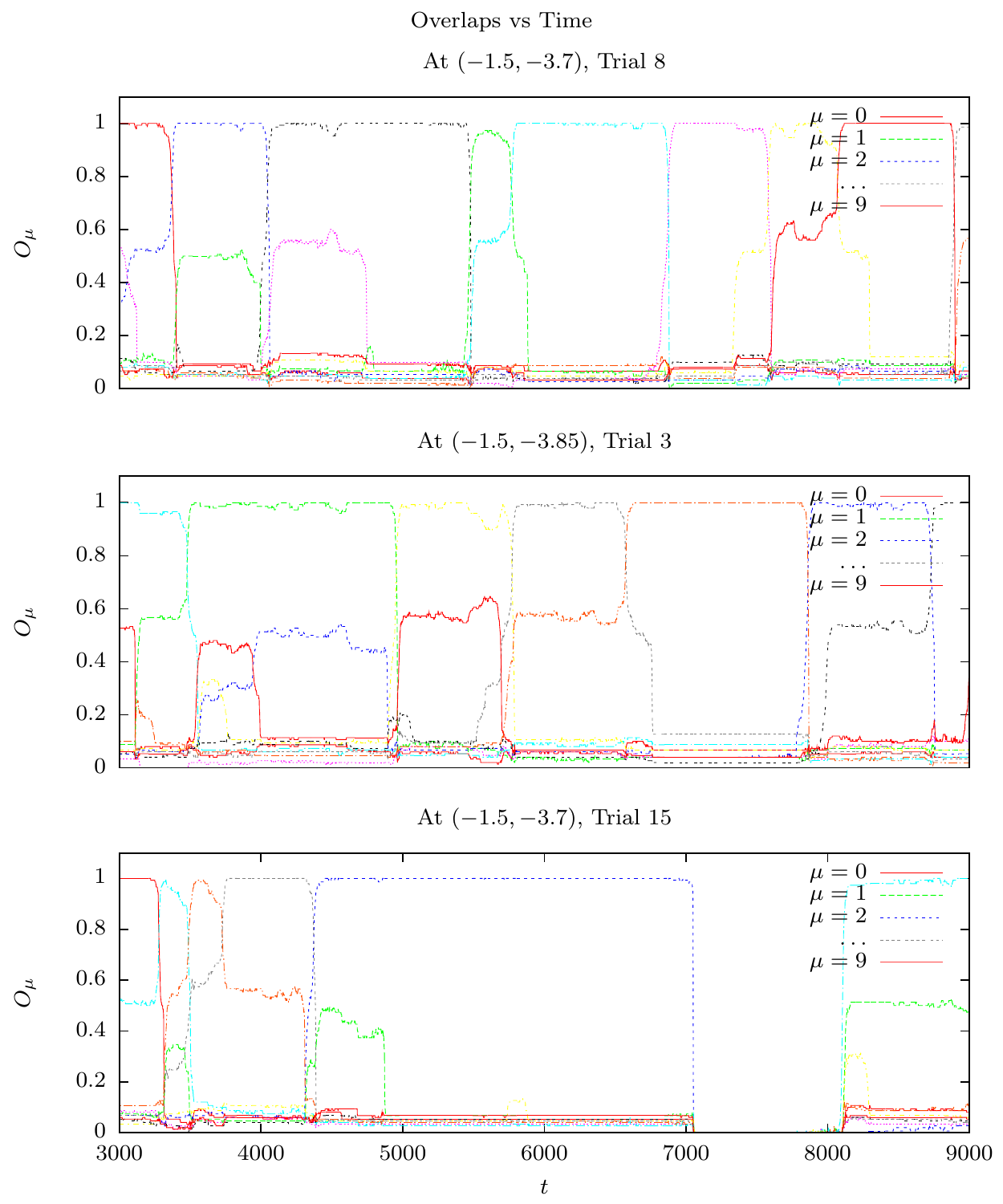}
\end{center}
\caption{Overlaps behavior over time at points where the error bars in figure~\ref{initsOE} change suddenly. Typical behaviors are shown in the top and middle panels. In the bottom panel, a null-out effect is observed. \label{errorbarProbes_O}}
\end{figure}

\begin{figure}
\begin{center}
\includegraphics{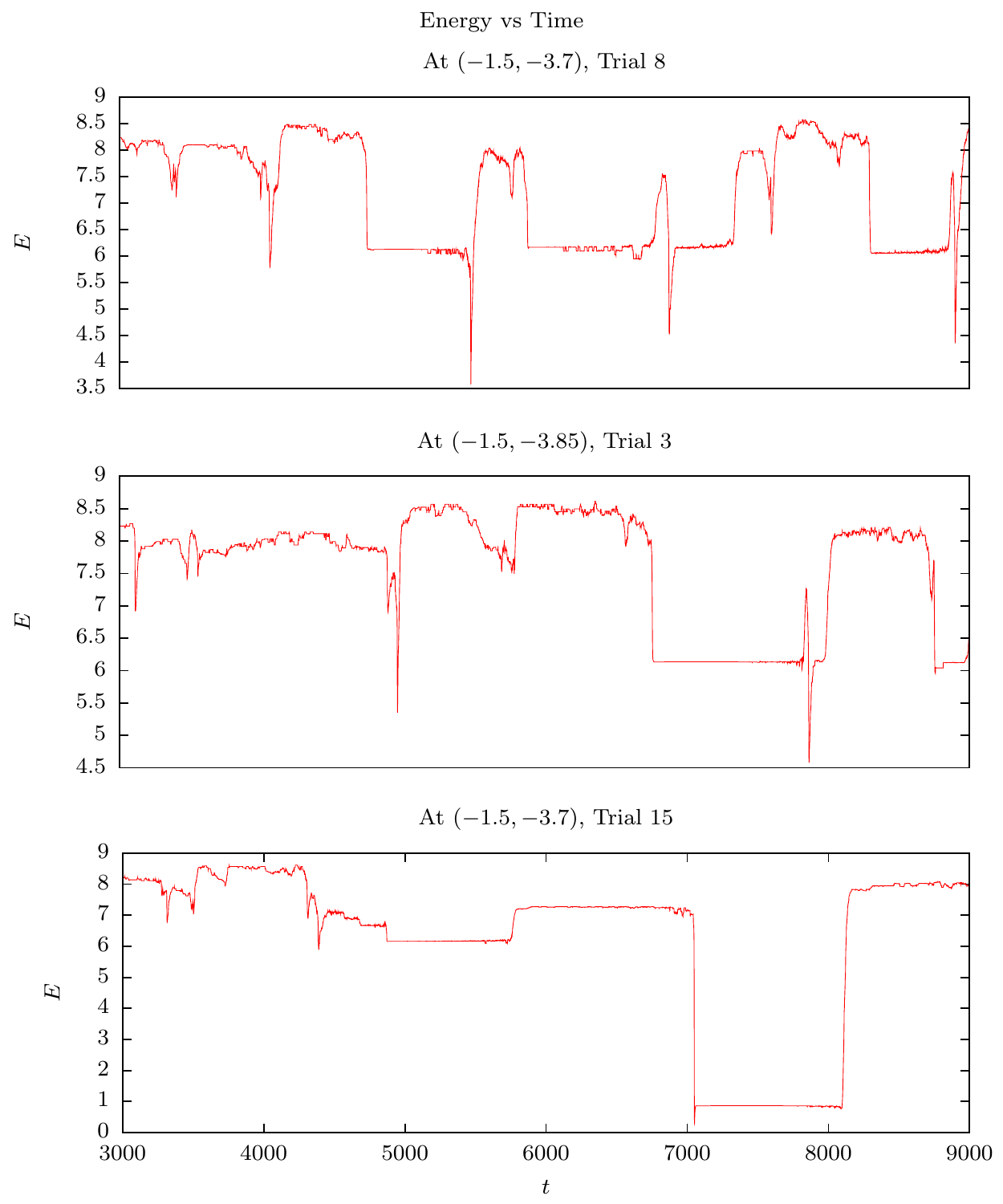}
\end{center}
\caption{Energy behavior over time at points where the error bars in figure~\ref{initsOE} change suddenly.  Typical behaviors are shown in the top and middle panels. In the bottom panel, a null-out effect is observed.  \label{errorbarProbes_E}}
\end{figure}

The behavior of the error bars in the lower panel of figure~\ref{initsOE} seems very  interesting. In our effort to understand the large variations in the error bars, especially the sudden change from $-3.7$ to $-3.85$,  we simulated again and examined our data for energies and overlaps at these points. The usual behavior of the system at these points is shown in figures~\ref{errorbarProbes_O} and \ref{errorbarProbes_E}, top and middle panels, for randomly selected trials (initial conditions). The apparent behavior of the graphs does not show much of a difference. However,  we noticed that the huge error bars are the result of few occurrences of a behavior that appeared in some trials, like in figures~\ref{errorbarProbes_O} and \ref{errorbarProbes_E}, bottom. It appears that the system virtually ``nulls out" sometimes. By the definition of overlaps, equation~(\ref{overlapsDef}), the null states are excluded from overlap calculation. So, the overlaps should vanish if all the units are in null states. To understand the behavior of the energy graph, notice that equations~(\ref{hamiltonian}) to (\ref{w}) tell us that the energies assigned to the null states are zero. However, the $-1$ values of $u_{sk}$ that appear in the sum in equation~(\ref{h}) make a nominal contribution to the total energy, making it slightly off zero. The system gets out of this ``resting state'' merely due to de-adaptation of other states and noise. The null-out did not occur in our trials at $-\log{\tau} \le 7.85$ in figure~\ref{initsOE}, hence minimal error bars. It occurs more frequently, and lasts for shorter periods of time as $-\log{\tau}$ gets larger, hence the decreasing error bars to the right.


Another interesting feature of figure~\ref{initsOE} is the difference between the energy and overlaps peaks, or rather bumps. While the rise in both graphs begins at about the same point on the right extreme of the panels for the reason that was explained before, the $\sigma_\mathrm{O}$ graph drops a bit later than the $\sigma_\mathrm{E}$ graph on the left. It is also much smoother than the $\sigma_\mathrm{E}$ graph. This behavior can again be understood by referring to figures~\ref{errorbarProbes_O} and \ref{errorbarProbes_E}. In these figures we see several transitions between patterns with very close energies. Such transition make up a significant portion of $\sigma_\mathrm{O}$, while in terms of energy, they mean little fluctuations.

\section{Behavior at around $\beta = 1$}\label{beta1}
\begin{figure}
\begin{center}
\subfloat[$\sigma_O$ for $-\log{\tau} = -2.25$]{\includegraphics[width=0.5\textwidth]{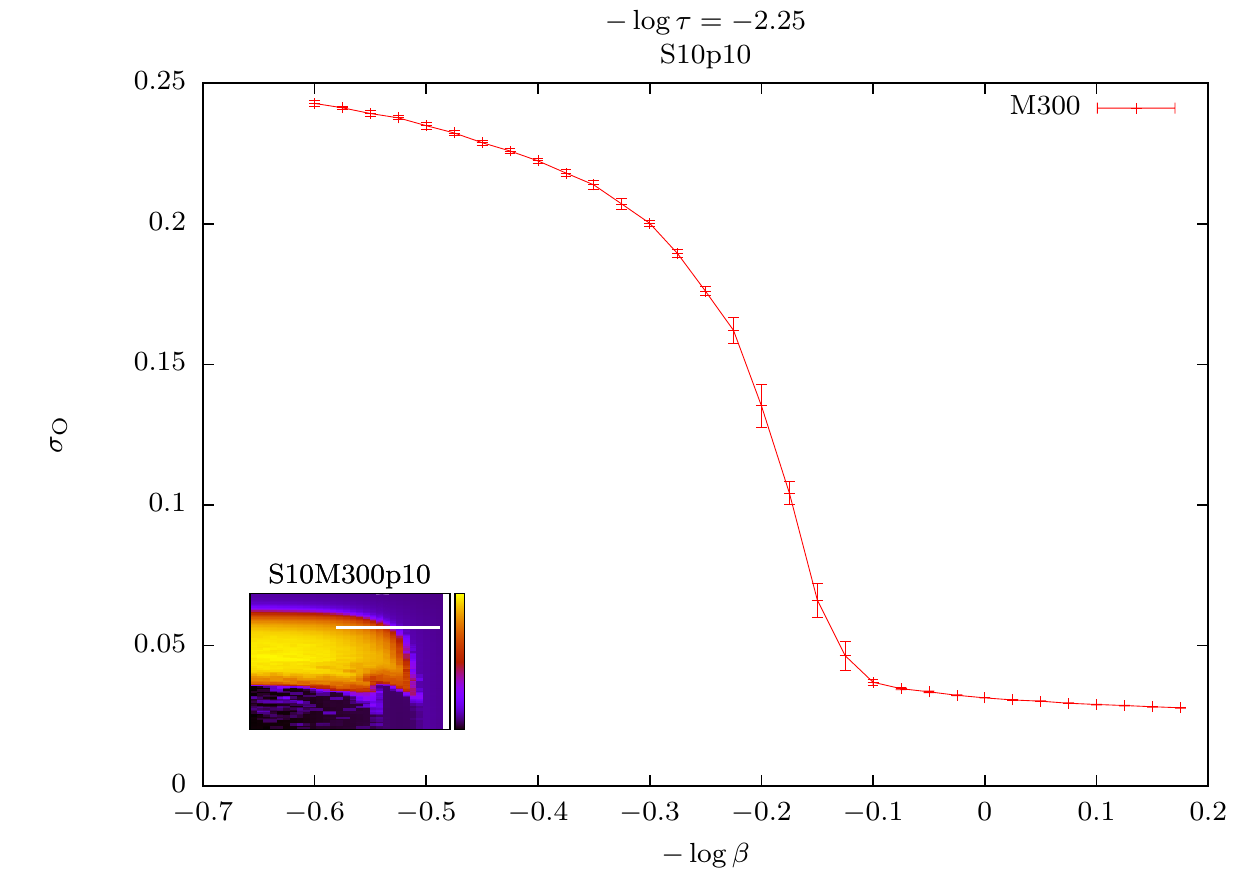}\label{t177_O}}
\subfloat[$\sigma_O$ for $-\log{\tau} = -4.65$]{\includegraphics[width=0.5\textwidth]{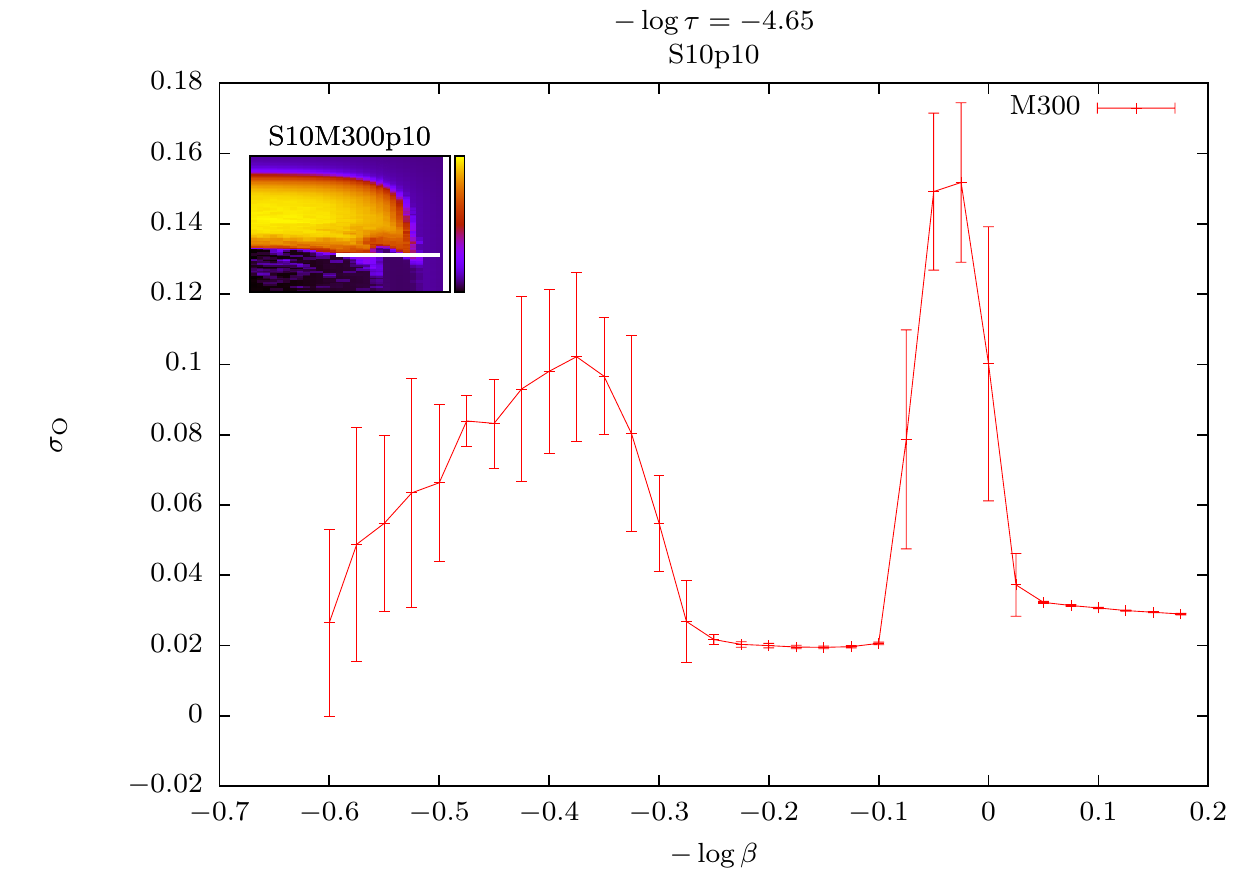}\label{t44668_O}}\\
\subfloat[$\sigma_E$ for $-\log{\tau} = -2.25$]{\includegraphics[width=0.5\textwidth]{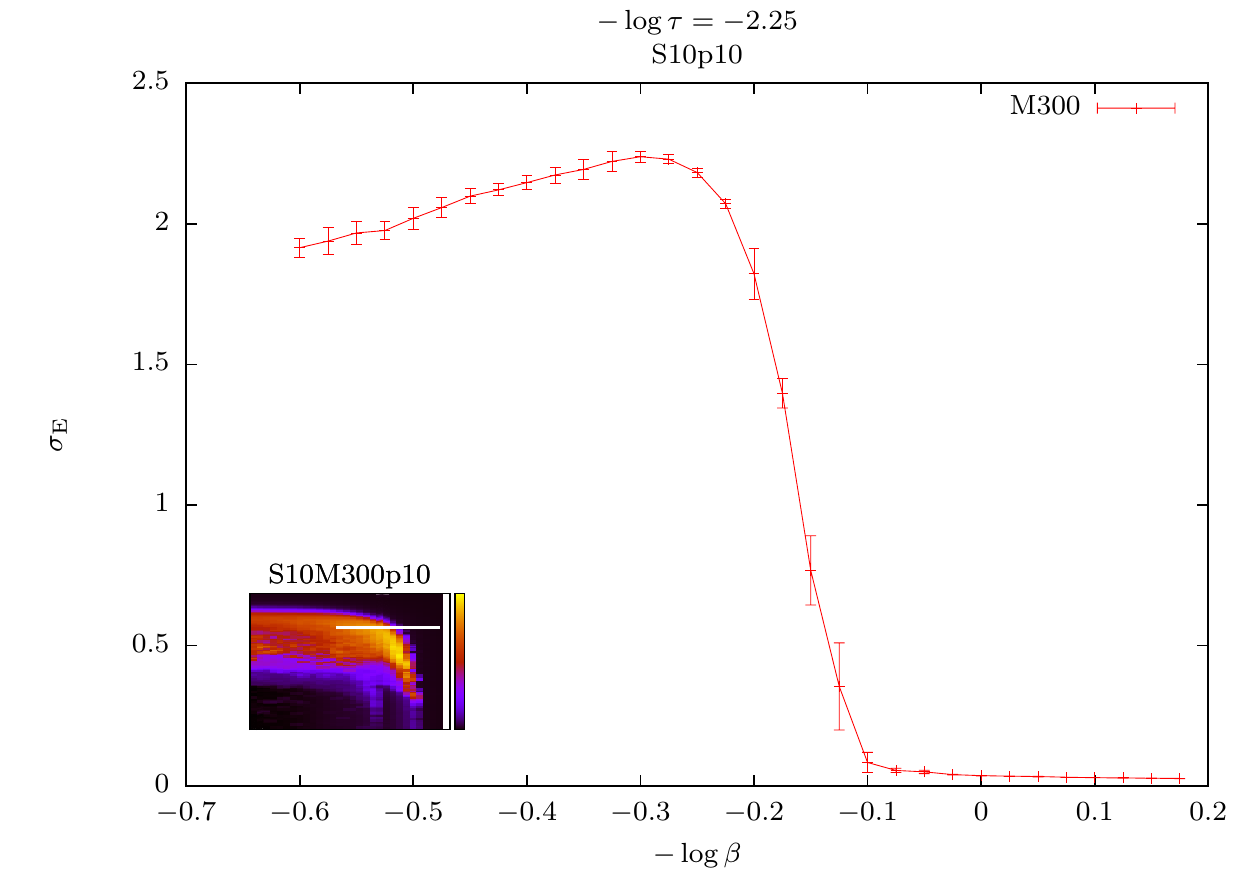}\label{t177_E}}
\subfloat[$\sigma_E$ for $-\log{\tau} = -4.65$]{\includegraphics[width=0.5\textwidth]{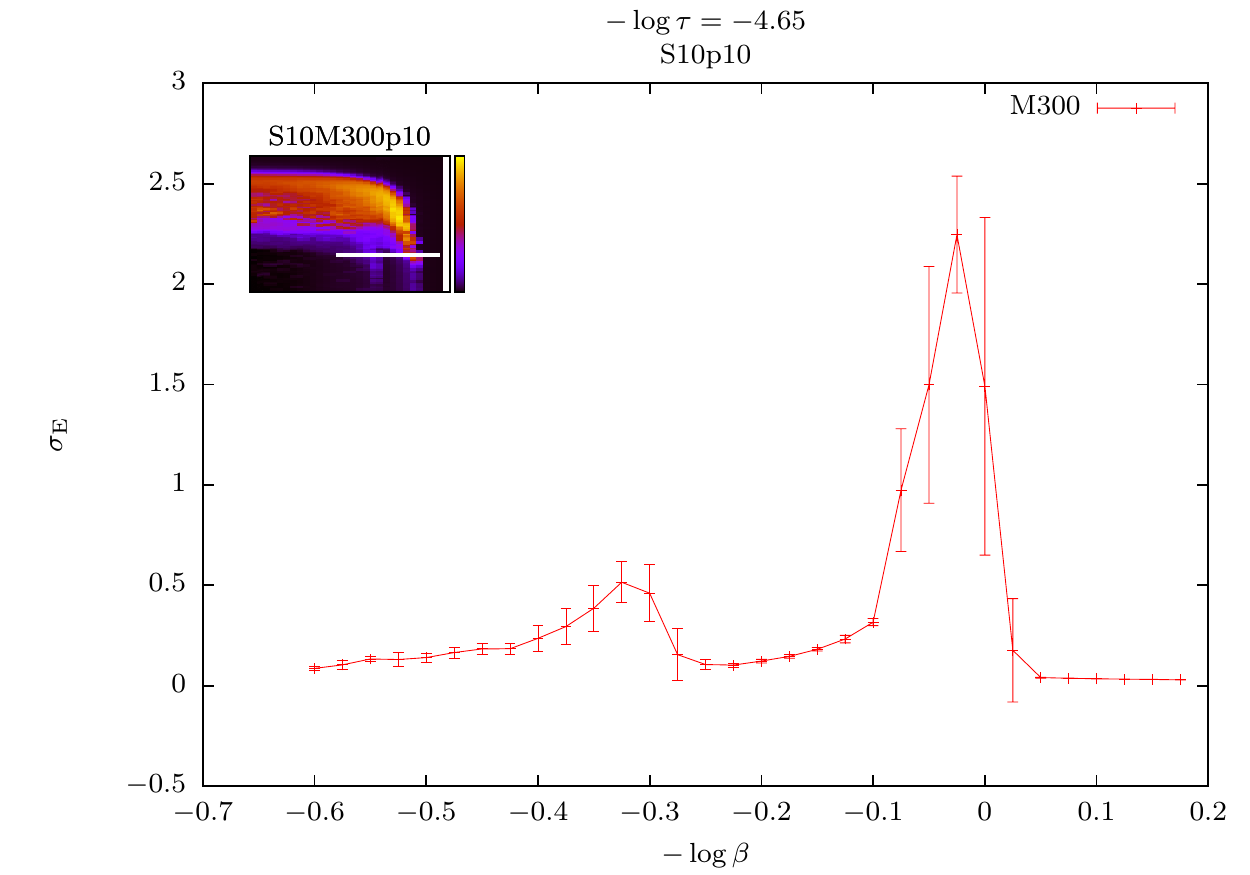}\label{t44668_E}}\\
\subfloat[$q_\mathrm{EA}$ for $-\log{\tau} = -2.25$]{\includegraphics[width=0.5\textwidth]{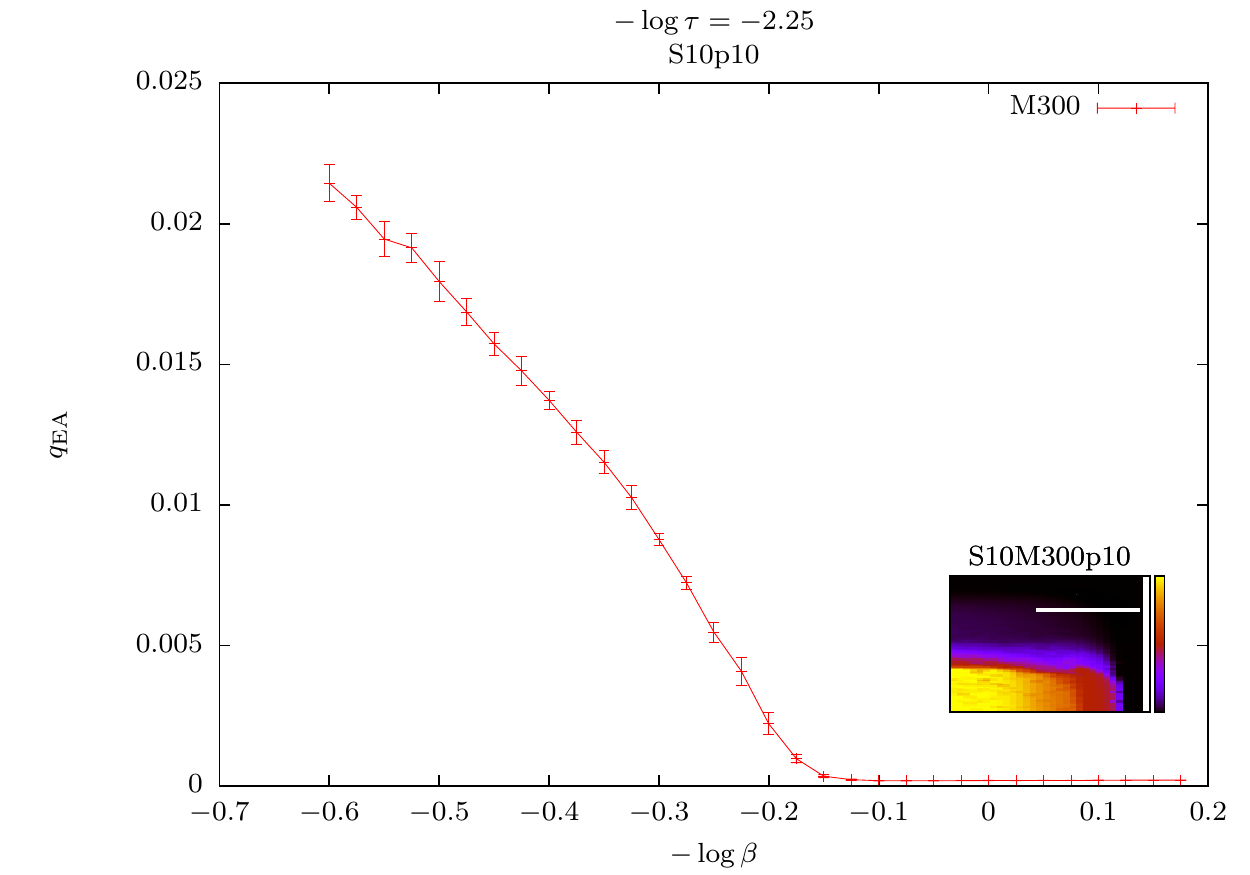}\label{t177_q}}
\subfloat[$q_\mathrm{EA}$ for $-\log{\tau} = -4.65$]{\includegraphics[width=0.5\textwidth]{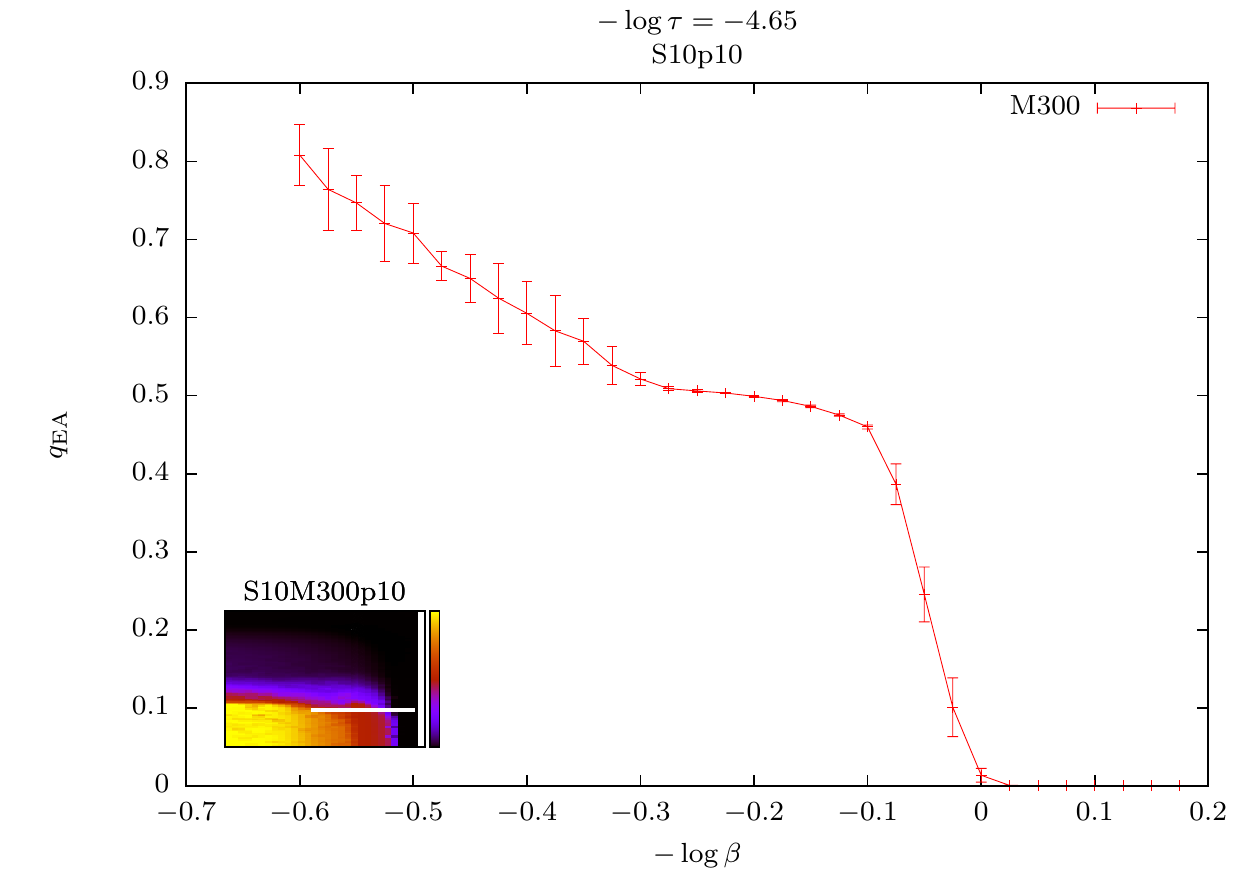}\label{t44668_q}}
\end{center}
\caption{Two horizontal sections of figure~\ref{longland} (insets). In the left panels, the run time is $9000$ steps, with the first $3000$ steps ignored. In the right panels, the run time is $5000$ with the first $1500$ steps ignored. All the inset graphs are from figure~\ref{longland} (left panels) with $5000$ steps run and $1500$ ignored. The right panels in the above figure completely match the inset graphs. The error bars here are calculated in the same fashion as figures~\ref{initsOE} and \ref{initsq}, i.e. 8 trials with different initial conditions/cue patterns.   \label{horizSections}}.
\end{figure}

Two different horizontal sections of figure~\ref{longland} (left panels) were selected for more detailed study in the region where noise has a considerable effect (figure~\ref{horizSections}). The section at $-\log{\tau} = -4.65$ is where adaptation is relatively slow, and the section at $-\log{\tau} = -2.25$ is where both noise and adaptation play a critical role in the behavior of the system (at around point $H$ in figure~\ref{longland}). More specific parameters are explained in the figure captions.

\begin{figure}
\begin{center}
\includegraphics{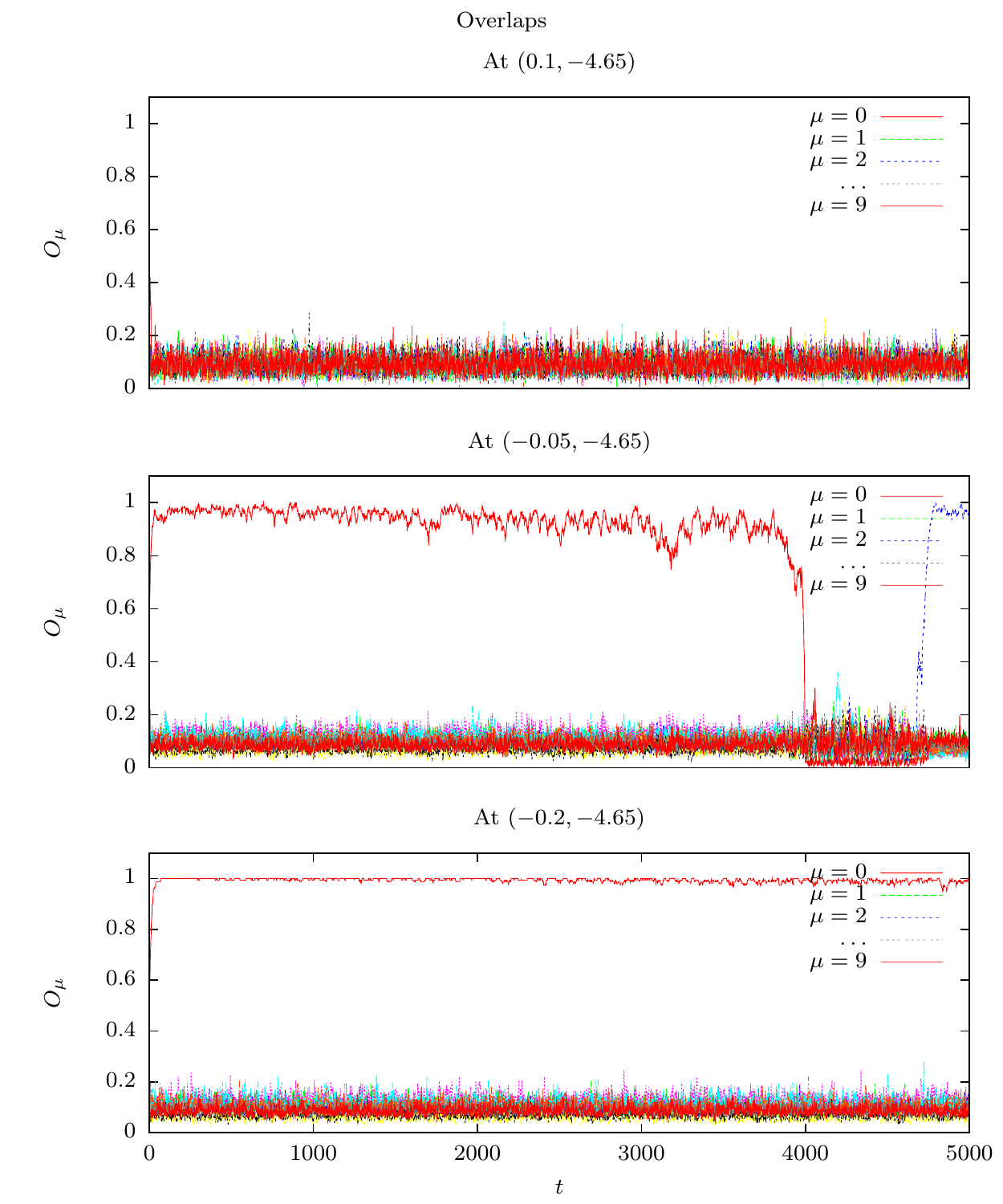}
\end{center}
\caption{Overlaps behavior at some select points in figure~\ref{horizSections} right panels. The top panel shows a dead/overactive dynamics. As $-\log{\beta}$ decreases, pattern retrieval begins (middle) and the retrieved patterns solidify as we further decrease the noise  (bottom).\label{b1probes_1}}
\end{figure}

\begin{figure}
\begin{center}
\includegraphics{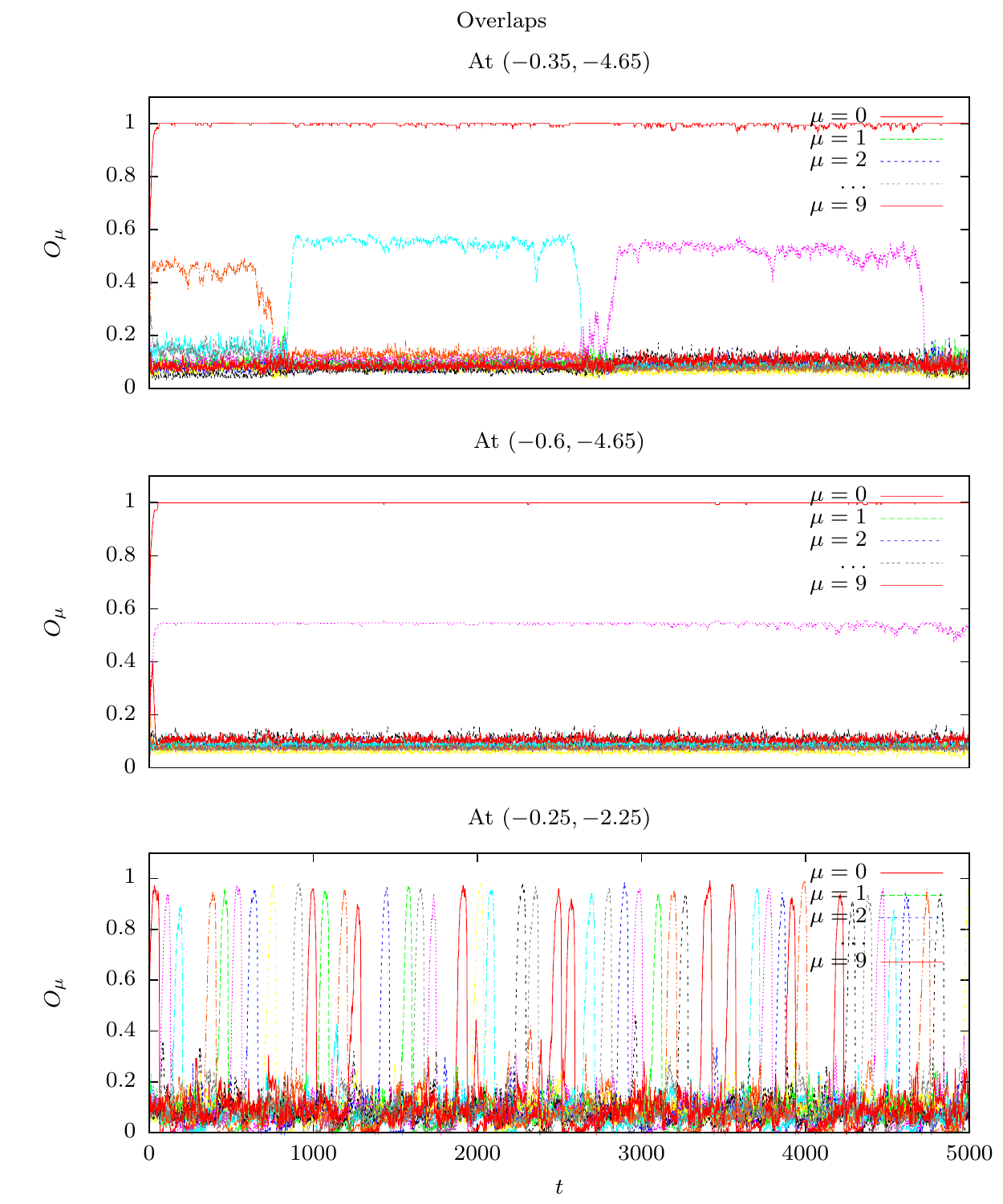}
\end{center}
\caption{Overlaps behavior at some select points in figure~\ref{horizSections}. The top panel shows a secondary pattern retrieval that occurs when noise is low enough. The primary and secondary retrieved patterns solidify as we further decrease the noise  (middle). In the bottom panel, the interplay between noise and adaptation is high, and the latching behavior is close to optimal. \label{b1probes_2}}
\end{figure}

To understand the behavior of the graphs in the right panels of figure~\ref{horizSections}, we choose to explain the overlap behavior in panel~\ref{t44668_O} as $-\log{\beta}$ decreases. At around $-\log{\beta}=0.1$ the noise is so strong that it does not allow any patterns to show up (figure~\ref{b1probes_1}, top).  A phase transition at around $\beta =1$ is a characteristic of an Ising models. A classical two-dimensional $q$-state Potts model also exhibits phase transition when $\exp({\beta})-1=\sqrt{q}$ \cite{Baxter73}.  The phase transition beginning at around $\beta = 1$ is not surprising. At around $-\log{\beta}=-0.05$  some jittering begins to show up (figure~\ref{b1probes_1}, middle). Notice that we are still close to the high-noise border, and the adaptation is slow but not zero, so it facilitates transitions induced by noise. This results in the first peak at around $0$. At around $-0.2$ the temperature is low enough for the retrieved patterns to stabilize (figure~\ref{b1probes_1}, bottom). Thus the overlap fluctuations decrease again at this point. However, a glance at the $q_\mathrm{EA}$ graph in panel~\ref{t44668_q} reveals that although a number of about $aM$($=0.5\cdot300$) units are fixed in the primary retrieved pattern, the rest of them are still fluctuating freely between various states. This can be seen better when the rest of the system is attracted to secondary patterns that are partially retrieved as shown in figure~\ref{b1probes_2}, top. This pattern retrieval needs a slightly lower temperature to occur, and the partially retrieved patterns are, like in the first peak, jittery and transient. This results in a smaller peak at around $-0.35$. As we further decrease the temperature, both partial and full pattern retrievals get solid and stable (figure~\ref{b1probes_2}, middle), resulting in low $\sigma_\mathrm{O}$ and $\sigma_\mathrm{E}$ values again. Here, we notice a ``life-shortening'' effect of noise on pattern retrievals. It can be observed in all of our data (including those not presented herein) that increasing noise alone results in a higher probability of pattern transitions, hence shorter retrieval lifetimes.


Now, we turn our attention to the other selected section, where $-\log{\tau}= -2.25$ (figure~\ref{horizSections}, left panels). Adaptation is faster in this section, so with decreasing temperature the first patterns show up in a lower temperature. The ascents in the $\sigma_\mathrm{O}$ and $q_\mathrm{EA}$ graphs look quite simple. The $\sigma_\mathrm{E}$ graph, however, shows an interesting peak immediately after the rise. Recalling that close-energy pattern transitions can account for overlaps activity with little energy fluctuations, we conclude that this peak signifies the most diverse pattern activity in terms of energy fluctuations. An instance of overlaps activity at $(-0.25, -2.34)$ is shown in figure~\ref{b1probes_2}, bottom. If we look back at figure~\ref{probesIBJ} we can see what happens if we further decrease the noise. We see that although pattern fluctuations may be fast due to fast adaptation, several patterns may rise at a time, that is, secondary and higher order pattern retrievals are observed in low temperatures. This ``purifying'' effect of noise was also observed in our study of the section $-\log{\tau} = -4.65$. Here, our results confirm that for a pure, distinct pattern retrieval we need $\beta$ close enough to~$1$.

We are now ready to articulate our optimality criterion and specify its region. We define a \emph{utility function} such that optimal latching corresponds to maximal utility function. Among various possiblities, we take our utility function $U(\beta, \tau, T)$ to be the number of \emph{transitions} between \emph{uniquely} retrieved patterns over a given rum time $T$. A pattern $\mu$ is ``uniquely'' retrieved when for some high and low thresholds $T_\mathrm{H}$ and $T_\mathrm{L} \in [0,1]$, we have $O_\mu > T_\mathrm{H}$ and $O_{\nu} < T_\mathrm{L}$ for all $\nu \neq \mu$. A ``transition'' occurs when a uniquely retrieved pattern is replaced by another. We calculate $U$ by counting the number of transitions.

The above utility function immediately excludes the dead/overactive region as not optimal. It also demands for the fastest latching dynamics. By ``fast'' we mean the life span of retrieved patterns and the transition time between retrievals are short. This requires that adaptation be maximal. The uniqueness condition for retrievals makes the noise maximal, too. With proper selection of $T_\mathrm{H}$ and $T_\mathrm{L}$, and a fixed $T$, the optimal region should get confined to around point $H$, or the bottom panel in figure~\ref{b1probes_2}.

\begin{figure}
\begin{center}
\includegraphics{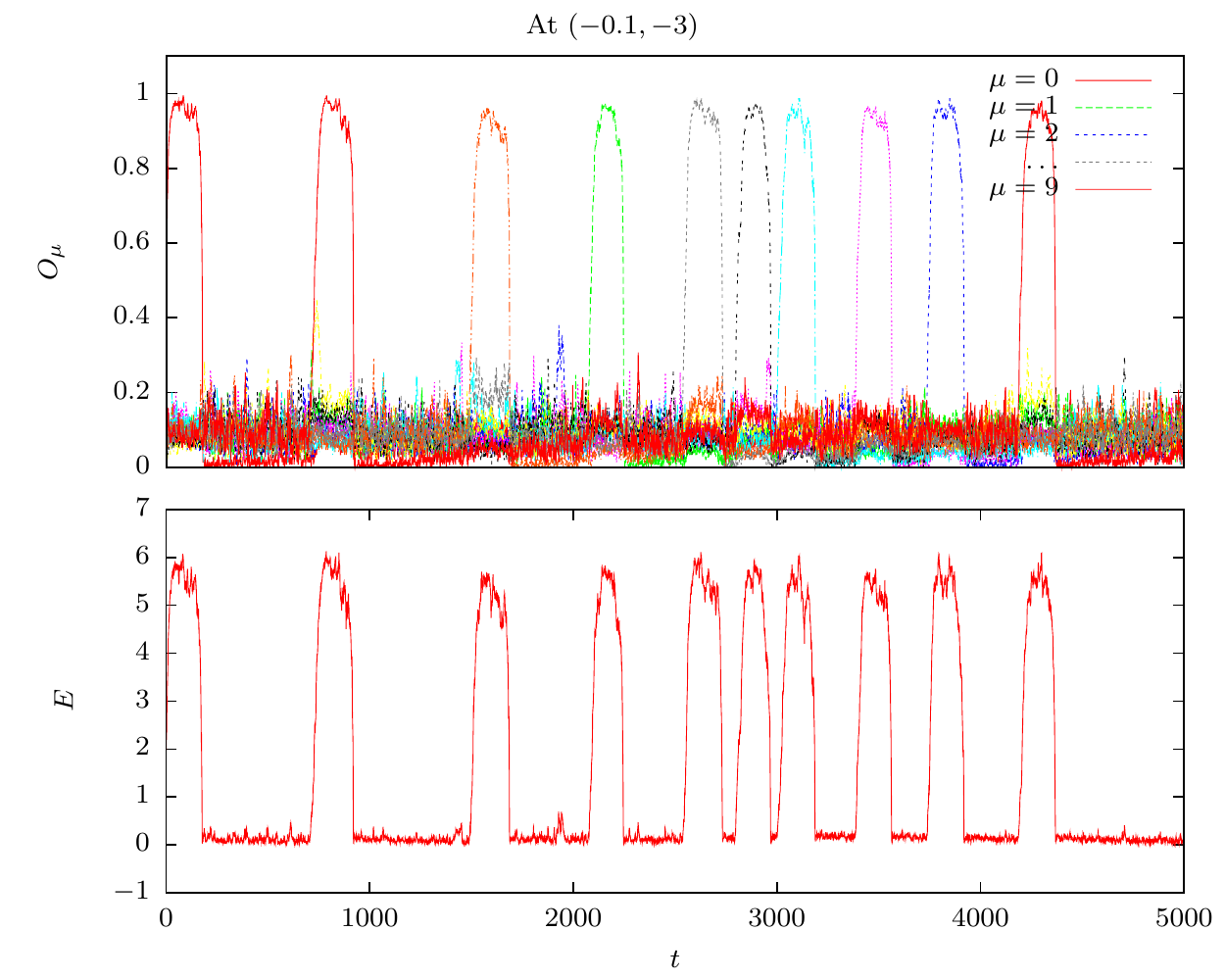}
\end{center}
\caption{Overlaps (top) and energy (bottom) behavior when $\sigma_\mathrm{E}$ is at maximum. Overactive periods separate the retrievals.\label{maxEnergyProbe}}
\end{figure}

Other optimality criteria are also possible to suggest. One can simply take $\sigma_\mathrm{E}$ as the utility function since it has an absolute maximum at around $(-0.1, -3)$ (cf figure~\ref{longland}). The overlaps and energy behavior at this point are plotted in figure~\ref{maxEnergyProbe}. Interestingly, we see that the retrieval periods are short, and the system spends considerable time in overactive/dead state (not null-out, compare with figures~\ref{errorbarProbes_O} and \ref{errorbarProbes_E}, bottom) during transitions where the average energy is almost zero.

As yet another option, one may look for the most ``diverse'' transitions as being optimal. By diverse, we mean having maximum randomness in terms of maximum entropy rate (assuming a stationary distribution):
\begin{equation*}
H = - \sum_{\mu \nu} p_\mu P_{\mu \nu} \log P_{\mu \nu}
\end{equation*}
where $p_\mu$ is the retrieval rate of pattern $\mu$, and $P_{\mu \nu}$ is the transition matrix. The search for this region shall be done in future studies.

\section{Discussion}
In this work, we have constructed a model combining two major characteristics from apparently separate disciplines. Our model possesses two major components: a temperature parameter and an adaptation one. The former is a primary constituent of a thermodynamical and statistical-physics framework, while the latter represents a major quality of real neural networks and plays an important role in the dynamics of realistic models suggested to date for the study of numerous phenomena in the brain. Figure~\ref{s10m300p10_OELand_top} reveals how these two basic components are joined to form a novel perspective - a latching behavior confined to a limited region of the parameter space.

A construction of Ising networks based on data from real retinal neurons suggests a preferred working temperature at around $\beta = 1$ \cite{Tkacik08}. A phase transition at this point is also observed in our model. The other basic parameter in our model, adaptation time-constant, also plays a key role in determining the type of network activity. The region where latching behavior occurs is limited in terms of noise and adaptation. However, more specific optimal criteria can be suggested to limit the desired area. The joint analysis of the two basic components, temperature and adaptation, singles out a critical region of optimal activity at around point $H$ in figure~\ref{s10m300p10_OELand_top}.  A comparison of the latching behavior at a sample point in this zone, such as $H$ (figure~\ref{probesAHD}, or \ref{b1probes_2} bottm), with several other points of latching possibility, such as $I$, $B$ or $J$ (figure~\ref{probesIBJ}), reveals how  indeed the optimal region is privileged: the retrieval sequence at $H$ exhibits fast and pure emergence of distinct patterns with regular periods, in contrast to co-occurring retrievals and indistinct, irregular transitions at other sample points. The findings here suggest that in the realistic models that incorporate adaptation mechanism, the respective time constants and the amount of noise might need to be limited to permitted ranges that comply well with the overall functionality of the network.

A rich variety of dynamical states are observed in different regions of the phase diagram. From a grammatical point of view, a traditional latching behavior occurs when a retrieval is cued by its previous retrieval, like in figure~\ref{probesIBJ}. However, with a sufficient presence of noise in the system, the network tends toward a spontaneous activity in which pattern retrievals are more or less cued by noise. This is most noticeable in figure~\ref{maxEnergyProbe}. In cases like point $H$, figure~\ref{probesAHD} middle, the transitions are highly noise driven, though the chain is not totally memory-less given the exponential recovery of adapted unit-states. Hence a deeper understanding of the boundaries and grammatical characteristics of these two types of behavior is definitely needed in future works.

Moreover, there are two types of dynamical states observed so far that can separate retrieval chains: overactive states (figure~\ref{maxEnergyProbe}) and null-outs (figure~\ref{errorbarProbes_E} bottom). The former is typical of the overactive/dead region where unit-states fluctuate too rapidly to form patterns. The latter occurs when noise is low enough for units to settle in null states, when the system is ``tired'' of recently retrieved patterns. This is, however, not a favorable state compared to pattern energy levels, hence it is a temporary state even though the null states do not adapt. Further work is required to verify these speculations and determine the rate and lifetime of such states.

Another interesting dynamical state is the `hierarchical' pattern retrieval exemplified in figures~\ref{b1probes_2} top, and \ref{probesIBJ}. In this sort of dynamics, ``one state is retrieved, serving as a framework for other states to be partially retrieved one after the other in the meantime,'' as described by a referee for this article. This as well seems very promising in terms of grammatical significance. Though further analysis falls out of the context of this article and remains for future studies.

As shown in section~\ref{beta1}, noise has a shortening effect on retrieval lifetime, or, an increasing effect on the rate of transitions.  In fact, noise is an essential constituent of the dynamics and unit-state transitions stop shortly as $\beta \to \infty$ (cf equation~(\ref{dynamics})). This accords well with the recent models \cite{Moreno-Bote07} in which alternations in dominant patterns of neural activity is induced by noise, while adaptation would not lead to alternations in the absence of noise. What is important in this scenario is that instead of an ad hoc assumption about the presence of noise, it is the interplay between adaptation and noise which sets the timescale of alternations. The fact is that the transition probabilities between different attractor states need not be at the scale of biophysical noise source characterized by fast timescales. This, indeed, would be too unrealistic given that the latching state of the network is meant to support transition states corresponding to highest cognitive states.  In terms of the state-space and energy landscape, the noise-adaptation interplay will shift the boundary line between basin of attractors  as well as reducing the depth of the minimum associated with dominant patterns \cite{Moreno-Bote07}. Given the optimal region in the noise-adaptation state space for maximum rate of transition probabilities there is room for realistic rate of alternations by varying noise and adaptation rates in the appropriate domain. In a similar vain, Kumar \etal \cite{Kumar10} have emphasized the rate of noise in shifting the dynamics in favor of spiking activity propagation in neural networks. The idea of a feed-forward network embedded in a recurrent network and hence the possibility of alternating patterns of activity in the form of a packet of synchronous neural activity bears a close resemblance to the hopping behavior of different attractor states in the Potts model. It will be interesting to see how the noise-adaptation interplay may play a similar role in controlling different activity modes in such embedded feed-forward networks.

The ``Potts'' virtue of this model, which lies in the multiplicity of states of each unit, plays a dramatic role in determining the shape and extent of latching region(s). The parameter $S$ was kept to be $10$ throughout this study. However, the effect of its alteration remains to be a target of future studies. Moreover, a thorough analysis of transition structure in the retrieval sequence is required to illuminate the potentials of the network for grammatical association and sequence generation. Any such analysis shall be preferably performed around the optimal region where the retrievals are unique, with high signal-to-noise quality, and frequent enough.


\ack{}
The authors would like to thank Yasser Roudi for his insightful comments and critical assessment, and Mohammad Reza Razvan for helpful suggestions at the early stage of this work. We also appreciate the critical comments and suggestions by the referees for this article, which spurred deeper analyses and new findings. The computation was carried out at Math. Computing Center of IPM (\url{http://math.ipm.ac.ir/mcc}).

\bibliography{biblio}


\end{document}